%% file: main.tex
\documentclass[conference]{IEEEtran}
\IEEEoverridecommandlockouts
\usepackage{cite}
\usepackage{amsmath,amssymb,amsfonts}
\usepackage{algorithm}
\usepackage{arevmath}     % For math symbols
\usepackage[noend]{algpseudocode}
\usepackage{graphicx}
\usepackage{textcomp}
\usepackage[normalem]{ulem}
\usepackage{xcolor}
\usepackage{fancyhdr}
\input{Macro}

\def\BibTeX{{\rm B\kern-.05em{\sc i\kern-.025em b}\kern-.08em
    T\kern-.1667em\lower.7ex\hbox{E}\kern-.125emX}}
\begin{document}

\title{ForensiBlock: A Provenance-Driven Blockchain Framework for Data Forensics and Auditability\\

% \thispagestyle{firstpage}

% \IEEEpubidadjcol

\thanks{"This work has been submitted to the IEEE for possible publication. Copyright may be transferred without notice, after which this version may no longer be accessible.”}
}

\author{
\IEEEauthorblockN{Asma Jodeiri Akbarfam\textsuperscript{1}, Mahdieh Heidaripour\textsuperscript{1}, Hoda Maleki\textsuperscript{1}, Gokila Dorai\textsuperscript{1}, and Gagan Agrawal\textsuperscript{2}}
\IEEEauthorblockA{\textsuperscript{1}\textit{School of Computer and Cyber Sciences, Augusta University}\
Augusta, USA\
\\{ajodeiriakbarfam, mheidaripour, hmaleki, gdorai}@augusta.edu}
\IEEEauthorblockA{\textsuperscript{2}\textit{School of Computing, University of Georgia}\
Athens, USA\
gagrawal@uga.edu}
}

\maketitle

\input{Sections/Abstract}


\maketitle
\input{Sections/Introduction}

\input{Sections/Background }

\input{Sections/Related-Work }
\input{Sections/Method }
\input{Sections/Queries}
\input{Sections/evaluation}
\input{Sections/Conclusion}

\bibliographystyle{IEEEtran}
\bibliography{bib} 

\end{document}

%% file: Macro.tex
\usepackage{xspace}
\usepackage{hyperref}
\usepackage[colorinlistoftodos]{todonotes}
\usepackage[linewidth=1pt]{mdframed}
\usepackage{mathtools}
\usepackage{enumitem}
\usepackage[makeroom]{cancel}
\usepackage{amssymb}
\usepackage{algorithm}
\usepackage {graphicx}
\usepackage{caption}
\usepackage{subcaption}
\usepackage{algpseudocode}
\usepackage{geometry}
\usepackage{pdflscape}
\usepackage{lscape}
\usepackage{csquotes}
\usepackage{tabularx}
\usepackage{color}
\usepackage{multirow}
\usepackage{tabu}
\usepackage{wrapfig}
\usepackage{soul}
\usepackage{ulem}
\usepackage{multicol}
\usepackage{float}

\usepackage{amsmath,amsfonts}

\usepackage{array}
\usepackage{textcomp}
\usepackage{stfloats}
\usepackage{url}
\usepackage{verbatim}
\usepackage{algpseudocode}
\newcommand{\mathmode}[1]{\ensuremath{#1}\xspace}
\newcommand{\tran}[1]{\mathmode{\mathtt{T}_{\text{#1}}}}

\newcommand{\access}{{\tran{AccReq}}\xspace}

\newcommand{\InitalUpload}{{\tran{InitialUpload}}\xspace}
\newcommand{\Case}{{\tran{FileUpload}}\xspace}

\newcommand{\Analysis}{{\tran{Analysis}}\xspace}

\newcommand{\AcessSC}{{\tran{AccessSC}}\xspace}

\newcommand{\Stage}{{\tran{Stage}}\xspace}
\newcommand{\MR}{{\tran{Provenance}}\xspace}

%% file: Sections/Abstract.tex
\begin{abstract}
Maintaining accurate provenance records is paramount in digital forensics, as they underpin evidence credibility and integrity, addressing essential aspects like accountability and reproducibility.   
Blockchains have several properties that can address these requirements. Previous systems utilized public blockchains, i.e.,    treated blockchain as a black box,  and 
benefiting from the immutability property. However, the blockchain was 
accessible to everyone, giving rise to security concerns and moreover, 
efficient extraction of provenance faces challenges due to the enormous scale and complexity of digital data. 
This necessitates a tailored blockchain design for digital forensics. Our solution, Forensiblock has 
a novel design that automates investigation steps, ensures secure data access, traces data origins, preserves records, and expedites provenance extraction.
%Forensiblock captures all provenance information related to investigation cases, ensuring data integrity.
Forensiblock incorporates Role-Based Access Control with Staged Authorization (RBAC-SA) and a distributed Merkle root for case tracking. These features support authorized resource access with an efficient retrieval of provenance records.   Particularly, comparing  two methods for extracting provenance records -- off-chain storage retrieval with Merkle root verification and a brute-force search -- 
the off-chain method is significantly better, especially as the blockchain size and number of cases increase. We also found that our distributed Merkle root creation slightly increases smart contract processing time but significantly improves history access. 
%Using Forensiblock, transaction processing time varies between different operations due to specific features tailored for digital forensics. 
Overall, we show that  Forensiblock  offers  secure, efficient, and reliable handling of digital forensic data.
\end{abstract}

%% file: Sections/Introduction.tex
\section{Introduction}

Digital forensics is crucial in modern investigations, enabling law enforcement agencies and organizations to extract, analyze, and preserve digital evidence for legal proceedings \cite{li2019blockchain, akhtar2022using}. However, safeguarding evidence, particularly the digital evidence, throughout the entire investigative process   remains a primary challenge \cite{borse2021advantages}. The Chain of Custody (CoC) plays a vital role in maintaining the integrity and credibility of digital evidence in digital forensics. CoC involves meticulously recording every transaction related to digital forensic evidence and maintaining a comprehensive storage history since its creation \cite{borse2021advantages}. Traditional evidence collection and preservation methods often struggle to provide comprehensive and transparent provenance records, leading to challenges in establishing trust and accountability \cite{chopade2019digital}. %{\bf should add some references to substantial previous sentence}

A standardized approach to preserving CoC is essential, highlighting the significance of data provenance in digital forensics \cite{pourvahab2019digital}.
Data provenance is the ability to trace and authenticate any data artifact's origin, custody, and history. 
 Particularly,  provenance records play a critical role in ensuring the integrity and reliability of evidence and addressing issues of evidence tampering and data manipulation. 

In recent years, blockchain technology has emerged as a promising solution for addressing  data provenance needs 
across different applications, including digital forensics~\cite{dasaklis2021sok}. As an immutable and decentralized ledger, blockchain offers a tamper-proof and transparent framework for recording and verifying the flow of digital evidence \cite{liao2021blockchain}. By implementing a provenance-driven blockchain system, it becomes possible to establish a robust and auditable  CoC for digital evidence, enhancing both privacy protection and data integrity.
%However, existing approaches often lack comprehensive automation and fail to incorporate essential elements, such as traceability, access control, provenance records, immutability, auditability, fast extraction with verification, and version tracking of data. This limitation arises from the prevalent treatment of blockchain as a black-box solution, primarily utilized for logging purposes in digital forensics workflows. Consequently, there is a pressing need for a design that specifically addresses the unique requirements of provenance in digital forensics while encompassing the mentioned capabilities. 

Several approaches have been designed to integrate blockchain into digital forensics%, such as digital forensic  CoC systems, and Criminal Investigation 
\cite{lone2018forensic, tsai2021application}. These approaches predominantly utilize public blockchains such as Ethereum. However, they face notable drawbacks that limit their widespread adoption and effectiveness. For instance, employing Ethereum as a public blockchain raises concerns about data accessibility and confidentiality, potentially jeopardizing sensitive investigative information \cite{ahmed2023using}. Furthermore, existing approaches in digital forensics on blockchains such as \cite{tasnim2018crab, ahmed2023using,pourvahab2019digital, li2019blockchain } have shown promising advancements in the domain. However, these approaches lack the integration of the essential elements,  which we argue are 
traceability, integrity \cite{bonomi2018b}, automated and secure authentication and access control,  immutability,  fast extraction of provenance records with verification, version tracking of data and secure communication between components. Particularly, in some cases, the limitations stem from the prevailing treatment of blockchain as a logging tool in digital forensics workflows, prioritizing alternative design aspects. 
%{\bf perhaps mention use of the off-the-shelf blockchains?}
Overall,  there is a  need for a framework based on a private blockchain that specifically addresses each 
of the above requirements of provenance in digital forensics. Such a design should solve privacy, security, and scalability challenges while preserving the desired benefits of transparency and immutability in digital forensics.
%In this context, the research questions emerge as follows: How can provenance-driven blockchain technology be effectively implemented in data forensics to address the complexities of safeguarding privacy, ensuring data integrity, and enhancing trust and accountability in the digital forensics landscape? Additionally, how can the limitations of existing blockchain systems for digital forensics be overcome to achieve enhanced evidence traceability, access control, provenance records, immutability, auditability, fast extraction with verification, and version tracking of data?

This paper presents ForensiBlock, a private blockchain solution specifically designed to overcome the limitations of existing blockchain systems in digital forensics provenance. ForensiBlock incorporates various essential components to enhance the investigation process and ensure comprehensive data management. It enables tracking all data involved in an investigation, including communication records, while providing the ability to trace data back to its origin. ForensiBlock also facilitates fast extraction and verification of evidence, ensuring efficiency and accuracy throughout the investigation. ForensiBlock  also utilizes cryptographic techniques to protect %communication and 
sensitive information. It includes access control mechanisms that detect and handle malicious access requests, ensuring that only authorized individuals have appropriate access to the data. The system also supports stage changes in the investigation, allowing for seamless progression and maintaining data integrity.

The main contributions of this paper are as follows:
\begin{itemize}
%\item Survey of provenance systems in various fields for a comprehensive overview
\item Identification of the significance of provenance in digital forensics and addressing the research gap in current blockchain-based digital forensics provenance research.

\item Proposal of a novel access control method specifically designed to meet the access control needs in digital forensics, automating the investigation process.

\item Presentation of a distributed Merkle tree for verifying the integrity of extracted cases.

\item Design and development of Forensiblock, a specialized 
framework tailored for digital forensics provenance that preserves all relevant records.

\item Implementation of extraction capabilities to enable timely retrieval of provenance records.

\item Conducting experiments to evaluate the capabilities and performance of Forensiblock.
\end{itemize}

The rest of this paper is structured as follows. Section \ref{section:Background} discusses an overview of blockchain, provenance, and digital forensics. Section \ref {section:Related Work} explores the related work. Section \ref{section:Framework} % discusses the importance of the method and 
describes the framework and the proposed protocols. Section \ref{section:Evaluation} 
illustrates the implantation and the experiment,  and the paper concludes with section \ref{section:Con}.

%% file: Sections/Background.tex
\section{Background} \label{section:Background} 

\subsection{Blockchain Technology} 

\begin{figure}[htbp]
\centering
\includegraphics[width=0.5\textwidth]{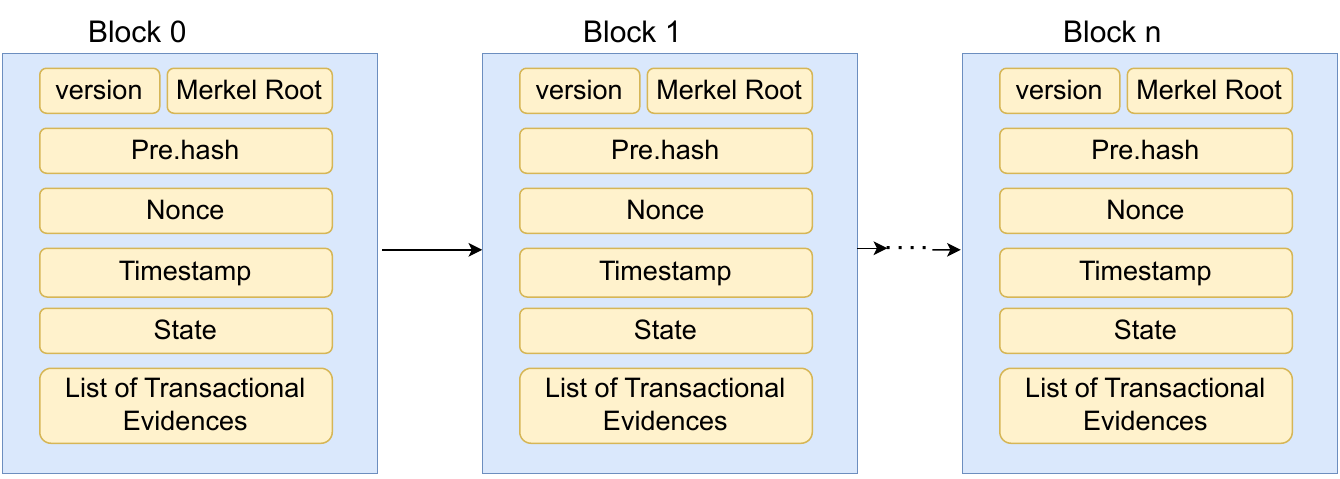}
\caption{chain of blocks}
\label{fig:chain}
\end{figure}

A  blockchain is a decentralized and distributed ledger that securely records transactions and stores information across multiple nodes in a network \cite{zhu2023blockchain, akbarfam2023dlacb, adhikari2023lockless}.  Key aspects of blockchain technology include mining, which is the fundamental mechanism employed by blockchain to secure the network through consensus algorithms \cite{nakamoto2008bitcoin}.  
%The basic 
functionality of blockchain provides a foundation for trust, transparency, and security in digital transactions.
%%Immutability is a key characteristic of blockchain that ensures data integrity and tamper resistance. The linking mechanism of blockchain, as described by Liao et al. \cite{liao2021blockchain}, involves creating a chain of linked blocks, as seen in Figure\ref{fig:chain}, where each block contains a cryptographic hash of the previous block \cite{bhutta2021survey}. This interdependence of blocks ensures that any modification made to a previous block would alter its hash, thereby invalidating all subsequent blocks. %This enhances the security and integrity of the overall blockchain system.
%% This notion of immutability enhances the integrity of the overall blockchain system.
Immutability is a critical characteristic of blockchain that ensures data integrity and tamper resistance. This property is achieved through two essential components: the storage of the Merkle root and the hash of the previous block. The Merkle tree, a significant data structure in blockchain technology, is vital in maintaining data integrity. It enables unique verification of data blocks without revealing other information \cite{jing2021review}. Additionally, the linking mechanism of blockchain, as described by Liao et al. \cite{liao2021blockchain}, involves creating a chain of linked blocks, as seen in Figure\ref{fig:chain}, where each block contains a cryptographic hash of the previous block \cite{bhutta2021survey}. This interdependence of blocks ensures that any modification made to a previous block would alter its hash, thereby invalidating all subsequent blocks. This notion of immutability, provided by the storage of the Merkle root and the hash of the previous block, enhances the integrity of the overall blockchain system.

%Key aspects of blockchain technology include mining, which is the fundamental mechanism employed by blockchain to secure the network through %proof-of-work 
%consensus algorithms \cite{nakamoto2008bitcoin}. 
%Blockchain can be categorized into different types; public and private. 
Blockchain technology encompasses various types, including public and private blockchains. Public blockchains, as exemplified by Bitcoin and Ethereum, are open to anyone. Conversely, private blockchains restrict access to a specific group of participants and are commonly used in enterprise settings for enhanced privacy and control \cite{amiri2021permissioned}. %Furthermore, the Merkle tree, a hash tree, is a significant data structure in blockchain technology. It ensures security and efficiency by allowing for unique verification of data blocks without revealing other information \cite{jing2021review}. 
 %Notably, some blockchains may also incorporate smart contracts. A smart contract is a self-executing program stored on the blockchain that enables decentralized decision-making and collaborative optimization \hoda{what does this mean? Is this correct? Smart contracts are executable programs stored on the blockchain and run when predetermined conditions are met. They offer programmability and flexibility in defining functions within the blockchain system.} in various industries, including manufacturing \cite{leng2023manuchain}. 
 Notably, some blockchains may also incorporate smart contracts. A smart contract is a self-executing computer program that operates on the Blockchain, automatically executing predefined actions according to specified conditions\cite{vu2023blockchain}.
These various aspects collectively contribute to blockchain technology's effectiveness and widespread adoption.

\subsection{Provenance}
Data provenance refers to understanding where data originates from and  its life cycle. It includes metadata describing an end product's origins, history, and evolution. Provenance, also known as lineage, encompasses a wide range of entities, data, processes, activities, and users involved in the entire process. With the exponential growth of digital data being created, copied, transferred, and manipulated through online platforms, provenance has become increasingly important in security. In the context of digital forensics, data provenance plays a critical role in establishing the legitimacy and origin of data, facilitating identification and reuse, and safeguarding the integrity of systems~\cite{abiodun2022data, pan2023data}.
%It offers assurance regarding the accuracy of data modifications, enables data forensics, and verifies access through a historical perspective 

\subsection{Digital Forensics}
Digital forensics deals  with retrieving and examining electronic device data. This process comprises five stages: 1) identification, 2) preservation, 3) collection, 4) analysis, and 5) reporting.
Potential evidence sources and individuals connected to the investigated device are determined during the identification stage. Preservation involves safeguarding all relevant electronically stored information (ESI) and documenting scene details. In the collection phase, digital information is gathered, creating duplicates for later analysis. The analysis stage involves a thorough search for evidence and its systematic examination. Finally, the reporting stage produces a comprehensive report following NIST guidelines~\cite{kent2006sp}.

\begin{figure}
\includegraphics[width=0.5\textwidth]{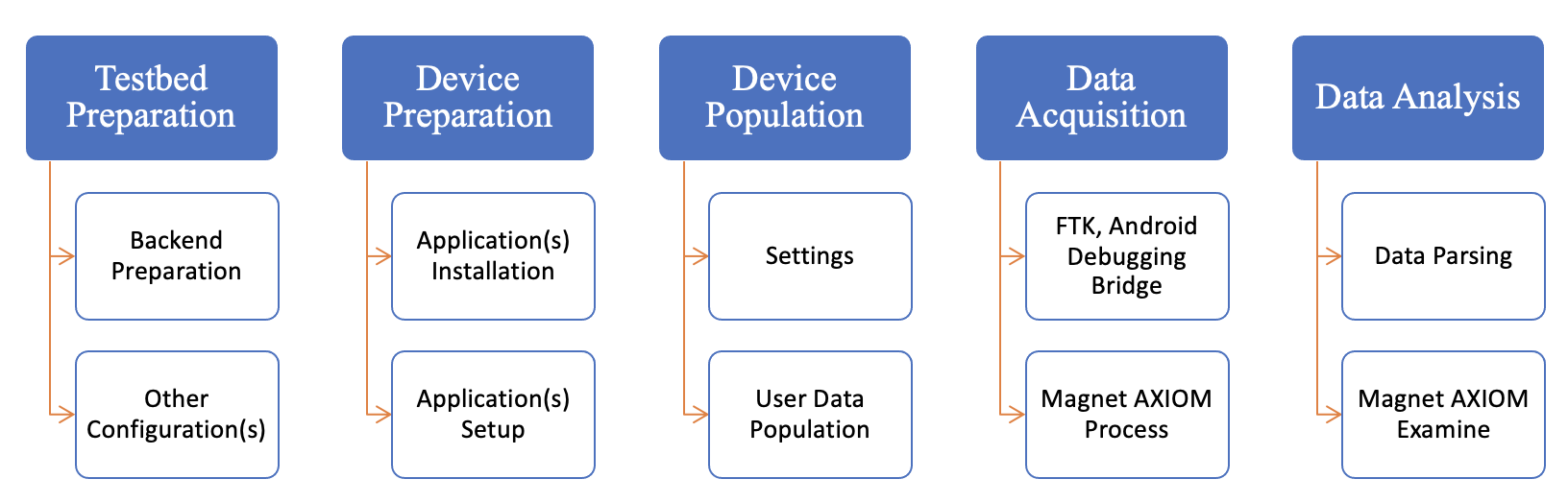}
\caption{Conceptual Overview of the Digital Forensics Process}
\label{dfc-fig}
\end{figure}

These five stages provide a systematic methodology for extracting, preserving, and analyzing digital evidence to maintain its integrity and admissibility in legal proceedings. Figure~\ref{dfc-fig} represents a conceptual overview of the digital forensics process used in our work.

%% file: Sections/Related-Work.tex
\section{Related work}  \label{section:Related Work}
% Provenance has been extensively studied in various domains, including copyright protection, intellectual property rights, scholarly contribution tracking, legal considerations, and digital forensics \cite{siddiqui2023activity}. In this section, we will investigate the benefits of blockchain technology to various domains in the context of provenance.

% Blockchain-based provenance mechanisms in IoT ensure integrity and verifiability by recording data operations in the blockchain network through transactions \cite{hu2020survey}. \cite{caro2018blockchain} developed AgriBlockIoT, a decentralized traceability system for Agri-Food supply chains, utilizing blockchain to record the entire supply chain and provide consumers with product history. Another work \cite{pahl2018architecture} combined IoT edge orchestrations with blockchain-based provenance to address trust concerns, recording origin and actions in the blockchain network. \cite{javaid2018blockpro} presented BlockPro, a secure IoT framework utilizing blockchain and physically unclonable functions (PUFs) for data provenance and integrity. Ali et al. proposed a secure provenance framework for cloud-centric IoT, incorporating blockchain to identify data origin and provide periodic traffic profiles \cite{ali2018secure}.

Blockchain technology has been extensively explored for recording data provenance in various domains, including  GDPR data collections\cite{neisse2017blockchain}, IoT (elaborated below),  supply chain management \cite{cui2019blockchain}, machine learning \cite{luthi2020distributed}, cloud computing \cite{zhang2017blockchain}, scientific workflows, legal  
scenarios and digital forensics \cite{siddiqui2023activity,troyer2021privacy, hoopes2022sciledger}. Systems  such as LineageChain \cite{ruan2021lineagechain} and BlockCloud \cite{tosh2019data} focus on detecting data modification attempts and implementing efficient query techniques and consensus protocols. ProvHL \cite{demichev2018approach} emphasizes access control management and user consent mechanisms. Duong and Dang \cite{dang2021effective} propose a public-permission model for open-access data. 
Provenance also holds significant importance in specific fields such as scientific workflows as well. In this context, works like BlockFlow \cite{coelho2019blockflow}, \cite{hoopes2022sciledger} SciLedger \cite{hoopes2022sciledger}, SmartProvenance \cite{ramachandran2018smartprovenance}, DataProv \cite{ramachandran2017using}, the work by Nizamuddin \textit{et al.} \cite{nizamuddin2019decentralized}, SciBlock \cite{fernando2019sciblock}, Bloxberg \cite{wittek2021blockchain}, and SciChain \cite{al2021scichain} introduce specialized approaches incorporating event listeners, voting systems, decentralized databases, timestamp-based invalidation, and unique provenance models. 
%{\bf give references to all areas and systems, for example, SciLedger is missing. } 
%However, generic provenance applications often lack specific data collection information. {\bf I don't understand the previous sentence}

%and private blockchains limit user privacy, validation mechanisms, and access to open data.
In recent years, the Internet of Things (IoT) has witnessed a remarkable expansion across various domains \cite{akbarfam2021proposing}. One noteworthy development in the IoT domain is incorporating blockchain-based provenance mechanisms.
In the domain of IoT, blockchain-based provenance mechanisms ensure integrity and verifiability through transaction records in the blockchain network \cite{hu2020survey}. Caro\textit{ et al.} \cite{caro2018blockchain} developed AgriBlockIoT, a decentralized traceability system for Agri-Food supply chains, utilizing blockchain to record the entire supply chain and provide consumers with product history. Pahl\textit{ et al.}\cite{pahl2018architecture} integrated IoT edge orchestrations with blockchain-based provenance to address trust concerns by recording origin and actions in the blockchain network. Javaid\textit{ et al.} \cite{javaid2018blockpro} presented BlockPro, a secure IoT framework utilizing blockchain and physically unclonable functions for data provenance and integrity. Ali\textit{ et al.} \cite{ali2018secure} proposed a secure provenance framework for cloud-centric IoT, incorporating blockchain to identify data origin and provide periodic traffic profiles. 
Provenance records are essential in digital forensics for preserving evidence integrity and authenticity. They establish data origin, prevent tampering, and ensure a reliable chain of custody. Authors of  \cite{li2019blockchain} introduced the IoT forensic chain (IoTFC), a forensic framework powered by blockchain technology, specifically designed to address the challenges of digital forensics in the Internet of Things (IoT) environment. These challenges include ensuring the trustworthiness of evidence items in digital forensics, maintaining continuous integrity checks for evidence items and examination events, providing hash validation for all evidence pieces. The framework emphasizes a comprehensive data provenance architecture and ensures the integrity of examination operations. However, this framework has several limitations, including: it ignores access control, which is a crucial component in managing the integrity and confidentiality of digital evidence. There is a lack of detailed communication between the various framework components, leading to some ambiguities, and the framework lacks evaluation for applicability and data extraction effectiveness.

%%In the paper \cite{borse2021advantages}, the authors introduced a protected file storage network based on peer-to-peer (p2p) architecture. This network utilized a decentralized Blockchain technology to store evidence, employing IPFS (InterPlanetary File System) for this purpose. The system's design, based on Hyperledger Sawtooth and a custom transaction family, ensures that only authorized personnel can access or maintain the evidence. Every transaction related to the documentation is recorded by the system right from the moment it is acquired.

Borse \textit {et al.} in \cite{borse2021advantages} introduces a novel approach: a hybrid blockchain solution that integrates features from both public and private blockchains. This solution is designed to effectively manage the Chain of Custody (CoC) in order to enhance transparency in evidence handling and transactions. While this method ensures the integrity and security of digital evidence in the field of digital forensics, it primarily focuses on maintaining the CoC within a specific investigative process. However, this approach does not consider access control or offer a comprehensive solution for data provenance in digital forensics.

 Ahmed \textit {et al.} \cite{ahmed2023using} proposes a Hyperledger based on a low-cost private blockchain and IPFS for media file tracking as evidence. The implemented access control allows the owner to access all capabilities for handling criminal records while other users have restricted access. However, the design lacks comprehensiveness, focusing on specific scenarios and a data forensic model. The access control model is simplistic, granting the data owner full access across various data forensic stages. Thus, there is a need for a more comprehensive solution that addresses all forensic authorization needs in data provenance and considers the stages of data forensics in the authorization component.

%%The paper\cite{agbedanu2023bloff} presents a blockchain-based forensic model for IoT called BLOFF. It utilizes the decentralized nature of blockchain to store and verify logs generated in IoT environments. The model focuses specifically on system and event logs. The entities in the model, including CSPs, Network Devices, and IoT devices, serve as blockchain nodes. New nodes are added through a key generation process, and their public keys are appended to transactions before being written onto blocks. CSPs act as miners due to their computational capacity. The model consists of a Blockchain Centre, Log Processing Centre, and User Centre.
Lone \textit{et al.} \cite{lone2018forensic} and Tsai \textit{et al.} \cite{tsai2021application} employed Ethereum, a public blockchain, to augment criminal investigations and establish a chain of custody mechanism. However, the utilization of public blockchains raises concerns regarding open-access and potential data privacy breaches
%\cite{lone2018forensic, tsai2021application} use the Ethereum Blockchain to enhance criminal investigations and implement a chain of custody system.
%The use of blockchains for provenance in digital forensics is often limited to leveraging the immutability feature without considering the specific design requirements for forensic investigations. Therefore, there is a need for a blockchain design tailored specifically for digital forensics. This design should automate investigation steps, ensure appropriate data access, trace data back to its source, maintain a comprehensive history, and enable timely extraction of provenance records. Introducing "Forensiblock," we propose a solution that addresses these requirements, offering significant improvements compared to previous approaches.

%%The utilization of blockchains for provenance in digital forensics often revolves around exploiting the immutability feature without considering the unique design requirements for forensic investigations or relying on public blockchains, which anyone can access. Consequently, there arises a need for a framework based on a private blockchain customized explicitly for digital forensics. This design should automate investigation steps, ensure proper data access, trace data origins, maintain a comprehensive history, and facilitate timely extraction of provenance records. In this context, we introduce "Forensiblock," a proposed solution that caters to these specific requirements, providing substantial advancements over previous approaches. 
Overall, the utilization of blockchains for provenance in digital forensics often revolves exploiting the immutability feature without considering the unique design requirements specific to the domain.
%{\bf since this is related work section, provide details of existing work and references}
Our work in this 
paper is distinct in offering a  private blockchain-based framework that is customized explicitly for digital forensics. 

%% file: Sections/Method.tex
\section{ForensiBlock Framework}%A Provenance-Driven Blockchain Framework for Data Forensics and Auditability} 
\label{section:Framework}

%As we have stated previously, Blockchain technology offers significant advantages in forensic applications, 
%specifically, such as tracking data versioning, managing access requests, and maintaining information provenance. By leveraging blockchain, transparency is improved throughout the investigation process, ensuring accurate data tracking, streamlined access request management, and maintaining data authenticity and integrity. This enhances the reliability and traceability of information.

\subsection{Motivation and Objective}

ForensiBlock addresses the following requirements for digital forensics:
\begin{enumerate}
\item Enhanced Evidence Traceability: ForensiBlock ensures enhanced evidence traceability by meticulously logging all steps in creating and modifying files. Additionally, individual tokens are generated for each piece of evidence, enabling seamless linking of items to their sources. 
\item Access Control Method: Our access control method caters to investigation stages and considers roles, granting appropriate access levels and enhancing data privacy. %Our novel access control method caters to varying investigation stages, granting appropriate access levels based on roles and stages and enhancing data privacy.

\item Provenance Records: ForensiBlock maintains detailed provenance records or system interactions,  promoting transparency and accountability during the investigation.

\item Immutability and Auditability: Leveraging blockchain technology, ForensiBlock ensures immutable and auditable evidence, meeting chain of custody requirements.

\item Fast Extraction with Verification:  ForensiBlock facilitates the swift retrieval of case-related information by securely storing it in the system's storage. Furthermore, it implements a robust verification method, ensuring the integrity, reliability, and accuracy of the accessed data.

\item Version Tracking of Data: ForensiBlock simplifies data version tracking and changes for investigators by linking versions together and maintaining a record, guaranteeing accuracy and reliability.
\end{enumerate}
In summary, ForensiBlock incorporates the entire investigation process, from case creation to data extraction, while ensuring secure access.

%Existing blockchain systems lack the comprehensiveness and organization of ForensiBlock. While they achieve evidence traceability, provenance records, and immutability, they lack structured data and design decisions tailored to digital forensics. {\bf add references here, also design decisions tailored  .. is a vague sentence, make it more 
%specific}

\subsection{ ForensiBlock Architecture}
\begin{figure}[htbp]
\centering
\includegraphics[width=0.5\textwidth]{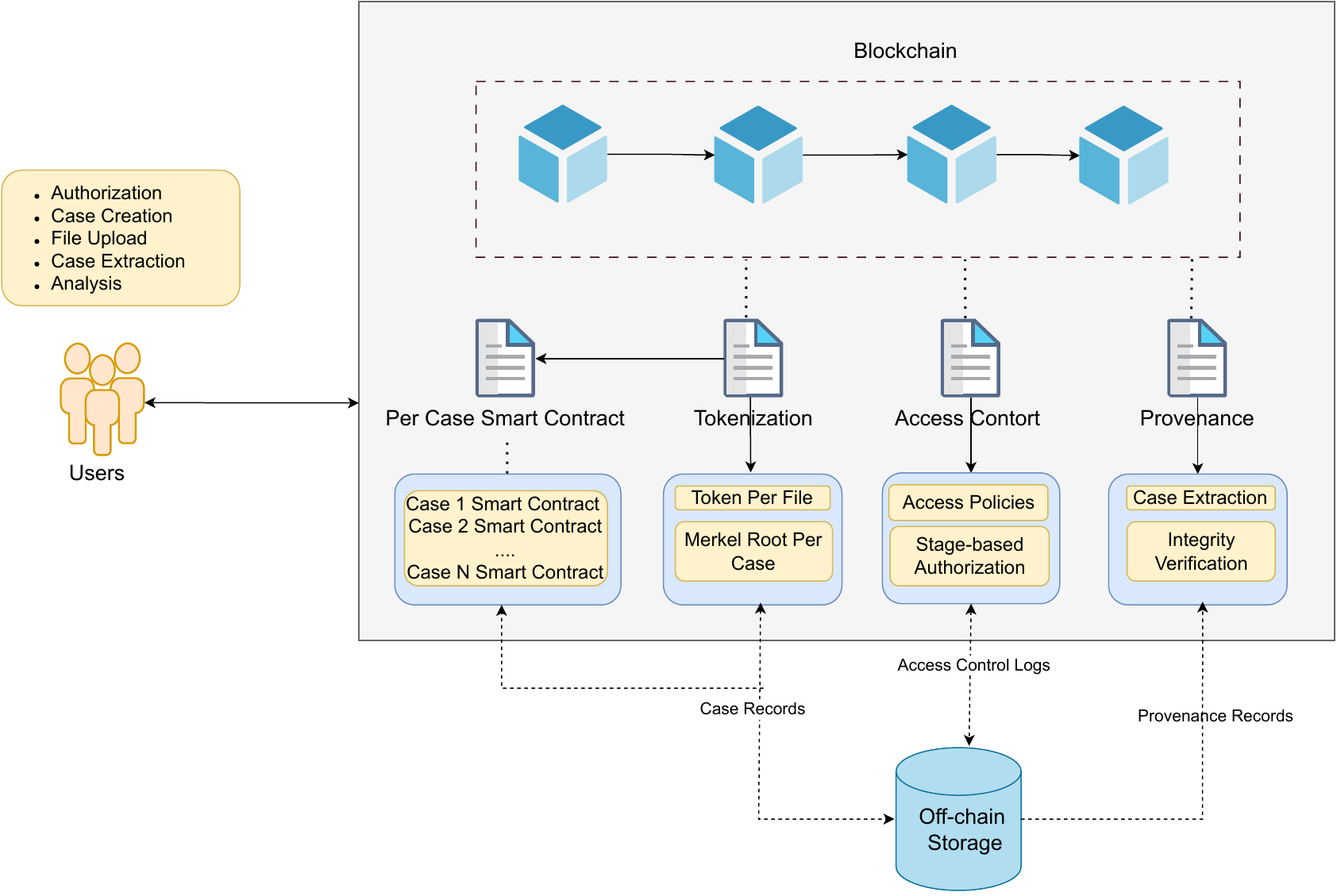}
\caption{ForensiBlock Architecture}
\label{fig:architecture}
\end{figure}

As illustrated in Figure \ref{fig:architecture}, ForensiBlock consists of three main %several 
components: blockchain, users, and storage. % that provide secure and traceable data storage, access control, and auditing capabilities. 
\setlength{\parindent}{0pt} 
\begin{itemize}
    \item Blockchain: Blockchain is used as the underlying technology to provide a decentralized and tamper-evident ledger for recording all transactions and data changes. 
    \item User Node:  The users of the data provenance platform in digital forensics are typically employees or authorized individuals associated with organizations or entities involved in digital forensic investigations.
    \item Off-chain Storage: The storage component is crucial in securely storing the digital forensic data associated with each case and maintaining the provenance recorded hierarchy for every case file. It receives encrypted data from users and ensures that authorized users can access the data based on the access rights determined by the access control smart contracts.
   Furthermore, the storage component maintains the provenance records for each case. This allows for faster records extraction and provides detailed data history, ensuring the stored information's integrity and traceability.
\end{itemize}

\subsection{Blockchain Structure}
%The blockchain ensures immutability and transparency, making it ideal for maintaining the integrity and provenance of digital forensic data.
ForensiBlock utilizes a customized private blockchain with four types of smart contracts that implement specific functionalities and streamline data management.
%\noindent 
%{\em Blockchain:} The underlying technology that provides a decentralized and tamper-evident ledger for recording all transactions and data changes. The blockchain ensures immutability and transparency, making it ideal for maintaining the integrity and provenance of digital forensic data.
%\begin{itemize}
 % \item   Blockchain: The underlying technology that provides a decentralized and tamper-evident ledger for recording all transactions and data changes. The blockchain ensures immutability and transparency, making it ideal for maintaining the integrity and provenance of digital forensic data.
    
\vspace{5pt}
\noindent \textit{Tokenized Smart Contract --} This smart contract is responsible for creating and managing cases within the blockchain. It also creates tokens for each file associated with a case. 
%A smart contract is generated for each case, recording metadata such as case number, timestamp, block number of the initial case, current stage, token list, and roles involved. The tokenized smart contracts also produce tokens upon request.

   \textit{Case Smart Contract--} This smart contract is created per each case and contains metadata such as case number, timestamp, the block number of the initial case, current stage, token list, and roles involved. The case-specific smart contract ensures isolation and modularity by containing the specific details of each case. This design enables the streamlined and organized administration of numerous cases, eliminating the necessity for a singular, all-encompassing contract to manage every case.
   %\hoda{ you can explain its purpose instead of saying what exact information it contains. If you have space you can do both. You can also indicate that this is generated by tokenized smart contract once a new case is create if you have not done it later on though}
    
   \textit{ Access Control Smart Contract--} It handles the system's access control mechanisms. This smart contract maintains a list of all cases and determines users' permissions and access rights based on their roles and the case stage. The access control smart contracts interact with the tokenized smart contracts to adjust access levels and validate user access requests.
   
  \textit{Provenance Smart Contract--} This smart contract facilitates the secure extraction of all related records associated with a case from the storage. Its primary objective is to ensure the integrity of the retrieved data by verifying that the case's records have not been maliciously tampered. % with or altered maliciously.
   The smart contract employs robust verification mechanisms to protect against potential data corruption originating from unreliable or malicious storage sources. The contract ensures that the extracted records remain trustworthy and free from unauthorized modifications by conducting thorough verification checks.
 % \subsection{User Node} 
 %  The users of the data provenance platform in digital forensics are typically employees or authorized individuals associated with organizations or entities involved in digital forensic investigations.
    
 %%  \subsection{Off-chain Storage} %The storage component is responsible for securely storing the digital forensic data associated with each case. It receives encrypted data from users and provides access to authorized users based on the access rights determined by the access control smart contracts. The storage component also stores encrypted keys for data access, ensuring data confidentiality.
  % The storage component is crucial in securely storing the digital forensic data associated with each case and maintaining the provenance recorded hierarchy for every case file. It receives encrypted data from users and ensures that authorized users can access the data based on the access rights determined by the access control smart contracts.
  % Furthermore, the storage component maintains the provenance records for each case. This allows for faster records extraction and provides detailed data history, ensuring the stored information's integrity and traceability.

\subsection{System Phases}
\begin{figure}[htbp]
  \centering
  \includegraphics[width=0.4\textwidth]{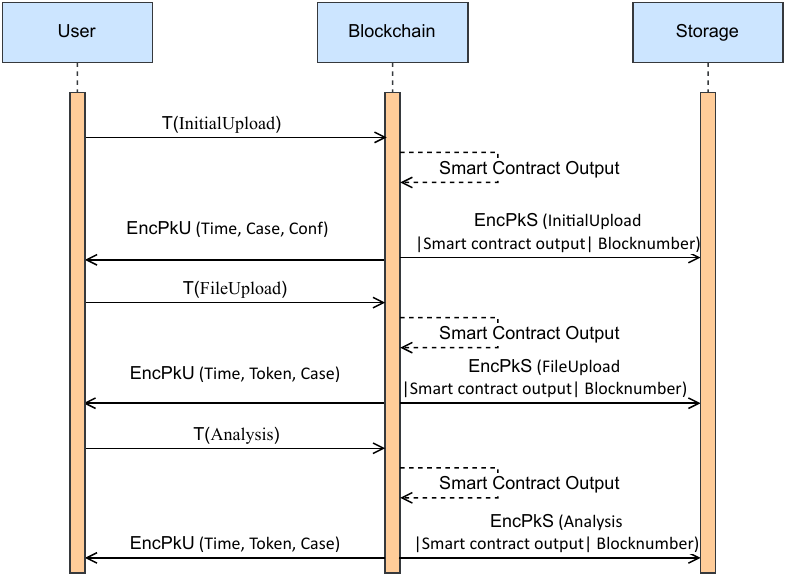}
  \caption{Transaction flow}
  \label{fig:upload}
\end{figure}

The data provenance platform operates through several phases that facilitate user interaction, data storage, retrieval, and analysis. These phases are as follows:
\begin{itemize}
    \item Phase 1, user registration:
    Similar to the registration process described in \cite{akbarfamdlacb}, the first phase of our system involves introducing users to the blockchain. This registration process is essential for establishing the user's identity within the system. To achieve this, the system adds the users' public key and performs a setup transaction to complete the registration process.
    \item Phase 2, uploading data and stage specification: As depicted in Figure \ref{fig:upload}, a case is created when users send a \InitalUpload transaction. This transaction contains the details of the case, including the case number. %to add a new case. 
    Once the blockchain receives this transaction, the tokenized smart contract is triggered, creating a new case smart contract specifically for the case. %, which includes important details such as the case number, timestamp, block number of the initial case, current stage, token list, and roles involved.
    Additionally, it communicates the case number and current stage to the access control smart contract through \AcessSC. Upon receiving \AcessSC, the access control smart contract adds the case number to the case list and sets permissions based on the case stage and the roles involved. %More details of the type of access control can be found in section \ref{AC}.
     % Fix the idea for this part
    % \hoda{the below details are confusing. Add some description so that I can edit the text. We have three smart contracts; when and how are they triggered? These details can be given here or when you explain each smart contract in the previous subsection.} 
    The tokenized smart contract's output, i.e., case smart contract, along with the \AcessSC, is logged on the blockchain. Simultaneously, the tokenized smart sends the encrypted transaction information and the output to the storage for provenance record keeping, while transferring the \textit{EncPkU(time, hashed data, case)} to the user.
    
    \item  Phase 3, uploading files: Once a case is created, the user can upload related files by sending a \Case per each file. 
    This triggers the tokenized smart contract, which generates a token for the file. 
    The token is then added to the list associated with the case number's smart contract.
    After logging on the blockchain, the tokenized smart contract forwards the token and the \Case transaction to the storage for provenance and sends the \textit{EncPkU (time, token, case)} to the user.
    
    % The  user sends data to the storage using EncPkS (token, data, case number). Upon receiving the data, the storage sends an approval transaction, \Ver, to the blockchain, ensuring a connection between the data and the token.
   % \hoda{at phase 4 you have not explained the data retrieval. Instead,  you have explained the change if stage. This should be clear in both the title and the details}
    %\asma{does it work now?}
    \item Phase 4, data access inquiry: Upon receiving \access from a user, the access smart contract 
    validates the request and assigns appropriate access levels. The \access and the user's access levels are encrypted and transmitted to the storage. 
    
    \item Phase 5, uploading new edits of previously accessed data:
    As illustrated in Figure \ref{fig:upload}, after analyzing the data, the user can send an \Analysis to the blockchain. This action triggers the tokenized smart contract to generate a token for the new file, add it to the list of tokens within the smart contract, and update the relevant case details. The subsequent steps in this phase are similar to phase 2.
      \end{itemize} 
      Note that, at any phase, a user with specified access privilege can submit a \Stage for changing the case stage. Upon receiving this transaction, the tokenized and access control smart contracts will change the case stage and adjust the rules accordingly. The system also considers the Extraction of provenance records which a user can extract the case records online or offline. To extract it online, the user must submit a \MR transaction. %     A case's records can be tracked offline or through \MR transactions. 
     Upon receiving \MR, the provenance smart contract will be provoked and returns all the provenance information related to that specific case.%If a user submits this transaction, the provenance smart contract will be provoked, and the user will receive all the provenance information related to a specific case.

%\begin{figure}[htp]
%   \centering
  %  \begin{subfigure}{6cm}
  %      \centering
    %    \includegraphics[width=\linewidth]{Sections/Figures/InitialUpload2.jpg}
    %    \caption{Initial Upload of Data}
    %    \label{fig:upload}
    %\end{subfigure}
    %\hfill

    %\begin{figure}[htp]
%   \centering
  %  \begin{subfigure}{6cm}
  %      \centering
    %    \includegraphics[width=\linewidth]{Sections/Figures/ new-trasnaction.png}
    %    \caption{Initial Upload of Data}
    %    \label{fig:upload}
    %\end{subfigure}
    %\hfill
 %   \begin{subfigure}{6cm}
     %   \centering
     %   \includegraphics[width=\linewidth]{Sections/Figures/Analysis2.jpg}
     %   \caption{Uploading New Edits}
        \label{fig:analysis}
  %  \end{subfigure}
   % \caption{Data Upload and Analysis}
 %   \label{fig:combined}
%\end{figure}
 % adding relevant transaction

\subsection {Protocols}
ForensiBlock introduces a set of protocols to enhance provenance records. It achieves this by recording every action, ensuring secure data access tailored for digital forensics, enabling efficient information extraction from dedicated storage, and implementing a robust integrity verification method. 
Detailed explanations of protocols used in Frensiblock are as follows.
\subsubsection{Role-Based Access Control with Staged Authorization (RBAC-SA)} \label{AC}
In digital forensics, available access control methods such as Role-Based \cite{sandhu1998role}, Attribute-Based \cite{hu2015attribute}, or Rule-Based \cite{afonin2016performance} approaches often have limitations due to unique access control requirements. Specifically, the need for permission changes at different investigation stages can lead to complications when employing Role-Based access control, including potential errors and ongoing challenges
%in defining and managing multiple permissions and policies 
\cite{akbarfamdlacb}. Attribute-Based access control can become complicated due to managing multiple attributes and policies related to evidence. Rule-Based systems may not be adaptable to the dynamic nature of investigation stages, leading to difficulties in updating rules. 
%{\bf give specifics of why attribute-based and rule-based won't be sufficient either}
%To enhance data log access and automate the access control process, ForensiBlock incorporates an access control smart contract. While various access control methods such as Role-Based \cite{sandhu1998role}, Attribute-Based \cite{hu2015attribute}, or Rule-Based \cite{afonin2016performance} approaches can be utilized, digital forensics has unique access control requirements compared to regular systems. There is a need for permission changes at different stages of an investigation, which can be challenging if Role-Based access control is employed. Editing the permissions associated with a role can potentially introduce errors and complications. Moreover, defining and storing multiple permissions and policies can lead to ongoing challenges, including the need to add more access continually, the inability to make new decisions, and significant memory consumption in the system \cite{akbarfamdlacb}.

To address these limitations, we propose a new protocol called Role-Based Access Control with Staged Authorization (RBAC-SA).  RBAC-SA combines role-based and rule-based access control, introducing stages for different investigation phases, i.e., Affidavit Warrant, Investigation, Analysis, Presented in Court, Judgement Day, Case Closed, and Potential Appeal. This approach enables different role permissions at each stage, adding a dynamic aspect to the access control framework.

%To address these limitations, we propose a novel method called Role-Based Access Control with Staged Authorization (RBAC-SA) to cater to access control needs in data forensics. RBAC-SA combines role-based and rule-based access control, incorporating various stages to enable different levels of permissions for roles at each stage, adding a dynamic aspect to the access control framework. The access control smart contract follows a rigorous authentication process to grant user access based on their public key and determine their resource access level. 
Our access control smart contract follows a rigorous authentication process, granting user access based on their public key and determining their resource access level. Investigation participants, such as Digital Forensics Examiner, Investigator, Legal Counsel, and Law Enforcement personnel, are assigned specific roles with predefined responsibilities and permissions, ensuring appropriate access privileges based on roles and enhancing security during the investigation.
As the investigation progresses, RBAC-SA dynamically adjusts permissions based on predefined rules. For example, only law enforcement and digital forensics examiners can access the evidence during Affidavit, expanding to other roles in later stages. This fine-grained control over data access by the progression of the investigation enhances security, prevents unauthorized access to sensitive information, and guarantees that only authorized individuals with a legitimate need can access and modify data during each investigation stage. Protocol \ref{algo:rbac-sa-algorithm} exemplifies the RBAC-SA mechanism, retrieving access information based on transaction details.

%RBAC assigns specific roles to investigation participants, such as Digital Forensics Examiner, Investigator, Legal Counsel, and Law Enforcement personnel, with predefined responsibilities and permissions. RBAC ensures appropriate access privileges based on roles, enhancing security and streamlining the investigation.

%RBAC-SA introduces stages for different investigation phases (Affidavit Warrant, Investigation, Analysis, Presented in Court, Judgement Day, Case Closed, and Potential Appeal), governing permissions and access rights for each role at each stage. As the investigation progresses, RBAC-SA dynamically adjusts permissions based on predefined rules. For example, only law enforcement and digital forensics examiners can access the evidence during Affidavit, expanding to other roles in later stages.
%By implementing differentiated permissions for roles at different stages, our RBAC-SA access control mechanism offers fine-grained control over data access by the progression of the investigation. This approach heightens security, prevents unauthorized access to sensitive information, and guarantees that only authorized individuals with a legitimate need can access and modify the data during each investigation stage. Algorithm \ref{algo:rbac-sa-algorithm} exemplifies the RBAC-SA mechanism, retrieving access information based on transaction details.
\begin{algorithm}
\floatname{algorithm}{Protocol}
\renewcommand{\thealgorithm}{1}
\caption{RBAC-SA Protocol}
\label{algo:rbac-sa-algorithm}
\begin{algorithmic}[1]
\Procedure{RetrieveAccessInfo}{transaction}
    \State Read role and stage information from files
    \State Stored\_stage = \textit{transaction['current\_stage']}
    \State Sender\_publik\_key = \textit{transaction['sender\_public\_key']}
    \If{ \textit{current\_stage} is invalid or $\neq$ Stored\_stage}
        \State \textbf{return} 'Invalid stage'
    \EndIf
    \State Retrieve \textit{public\_key\_role} based on Sender\_public\_key
    \If{\textit{public\_key\_role is None}}
        \State \textbf{return} 'Access Denied'
    \EndIf
    \State Retrieve \textit{public\_key\_rights} based on \textit{current\_stage} and
    \State \textit{public\_key\_role}
    \If{\textit{public\_key\_rights is None}}
        \State \textbf{return} 'No access rights'
    \EndIf
    \State \textbf{return} \textit{public\_key\_rights}
\EndProcedure
\end{algorithmic}
\end{algorithm}

%% file: Sections/Queries.tex
\subsubsection{Provenance Records Management Protocol}%Provenance Records Protocol}
In designing ForensiBlock, our primary goal is to achieve effective storage and extraction of provenance records related to investigations. While blockchain storage suffices for some scenarios, efficient information is crucial for enabling effective analysis, facilitating seamless collaboration, and ensuring proper audit and accountability.
Forensiblock incorporates various provenance records, as depicted in Table \ref{table:provenance}. We have devised the Provenance Record Management Protocol to ensure the preservation of these records, track potential malicious modifications, and verify their authenticity. This protocol involves different strategies utilizing blockchain and off-chain storage, which will be further elaborated below.

\renewcommand{\arraystretch}{1.5}
\begin{table}[h]
\centering
\caption{Provenance Records for the ForensiBlock System}
\label{table:provenance}
\begin{tabular}{m{12mm}m{38mm}}
\multicolumn{1}{l}{\textbf{Record}} & \multicolumn{1}{l}{\textbf{\textbf{Description}}} \\
\hline
\multicolumn{1}{l}{Case Number} & Unique identifier for a case \\
\hline 
\multicolumn{1}{l}{Timestamp} & Time of the action/event \\
\hline
\multicolumn{1}{l}{Initial Block Number} & Block where the initial case \newline  transaction was recorded \\

\hline
\multicolumn{1}{l}{Current Stage} & Current stage of the case \\
\hline
\multicolumn{1}{l}{Token List} & List of tokens associated with \newline the case \\
\hline
\multicolumn{1}{l}{Access Request} & User request to access specific \newline case data \\
\hline
\multicolumn{1}{l}{Client Info} & Information about the user/client \\
\hline
\multicolumn{1}{l}{Token dependency} & analyzed Tokens derived \newline from a files \\
\hline
\multicolumn{1}{l}{Access Validity} & Matching system permissions \newline for the access request \\
\hline
\multicolumn{1}{l}{Stage Change} & Change in the tokenized smart \newline  contract's stage \\
\hline
\multicolumn{1}{l}{Type of Data Upload} & Indicates raw data or  \newline analyzed data \\
\end{tabular}
\end{table}

\begin{figure}[htbp]
  \centering
  \includegraphics[width=0.4\textwidth]{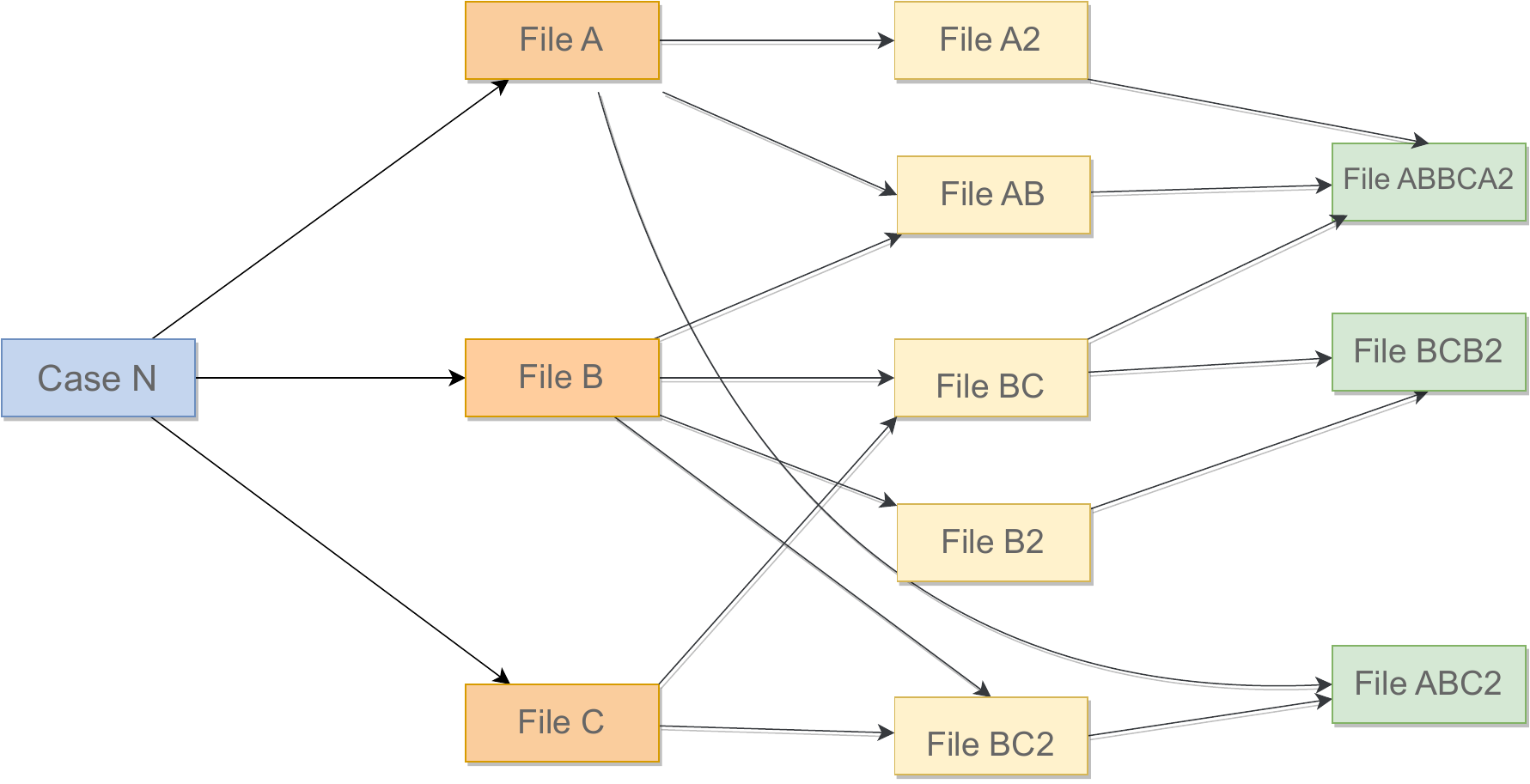}
  \caption{File dependency example per case}
  \label{fig:dep}
\end{figure}

\paragraph{File Version Tokens}%Tokens for Versioning of files for Provenance Tracking}
The relationship among our data exhibits complexity, as illustrated in Figure \ref{fig:dep}. The analyzed data can either be independent or dependent on one another. For instance, files A and B were combined to create file AB, while files B and C were utilized to generate file BC2. Subsequently, files A and BC2 were employed to produce ABC2.

To manage this data complexity, we propose a token generation algorithm that generates distinct tokens per file, acting as unique identifiers for file versions.
This algorithm creates tokens for two types of files: original files and derived files.
When an original file $i$ is added or modified, the token $K_i$  representing the file's version is generated. 
On the other hand, when a derived file is created based on $n$ initial files with tokens  $K_1, K_2, \ldots, K_n$, a derived token, $K_{\text{derived}}$, is generated. This token is created by hashing the tokens of the initial file, i.e., \[K_{\text{derived}} = \text{Hash}(K_1||k_2|| \ldots||  K_n,  \text{Time})\]

Here,  $||$ indicates concatenation, and $Time$ represents the timestamp of the corresponding transaction. Note that including the time in the hash calculation ensures that even if the same files are combined multiple times, distinct derived tokens will be generated due to the variation in timestamps.
This token creation method ensures that the token of the derived file is based on the tokens of its parent files.
By applying this algorithm, each case captures the versioning steps and relationships of the files associated with that case. 
%The Merkle DAG provides an organized and efficient means of tracking the provenance and version history of files within the blockchain. The tree will be stored at the off-blockchain storage.
%The integrity and consistency of files at each version can be verified using the Merkle DAG, as any change in a file will impact the hash values and the Merkle root of the DAG. This ensures secure and reliable provenance tracking within the blockchain system.

\paragraph{Enhancing Provenance Queries}
To enhance the extraction of provenance records, two methods are utilized. The first method involves storing the provenance records on the blockchain. %, which ensures their trustworthiness. 
The blockchain maintains additional information, such as the unique case number associated with each transaction. This facilitates the extraction of relevant transactions linked explicitly to the desired case. The design incorporates tokens as unique identifiers for each file within a case, which are stored in every case's smart contract. The transactions recorded in the blockchain include these token identifiers, establishing a seamless link between the data and the tokens. This mechanism simplifies the traversal of the blockchain. It enables the construction of a hierarchical representation, resembling a tree-like structure, that effectively captures the sequence of edits made to the data over time. The immutable and trustworthy nature of the blockchain ensures the reliability of the extracted provenance records.

\begin{figure*}[t]
  \centering
  \includegraphics[width=0.65\textwidth]{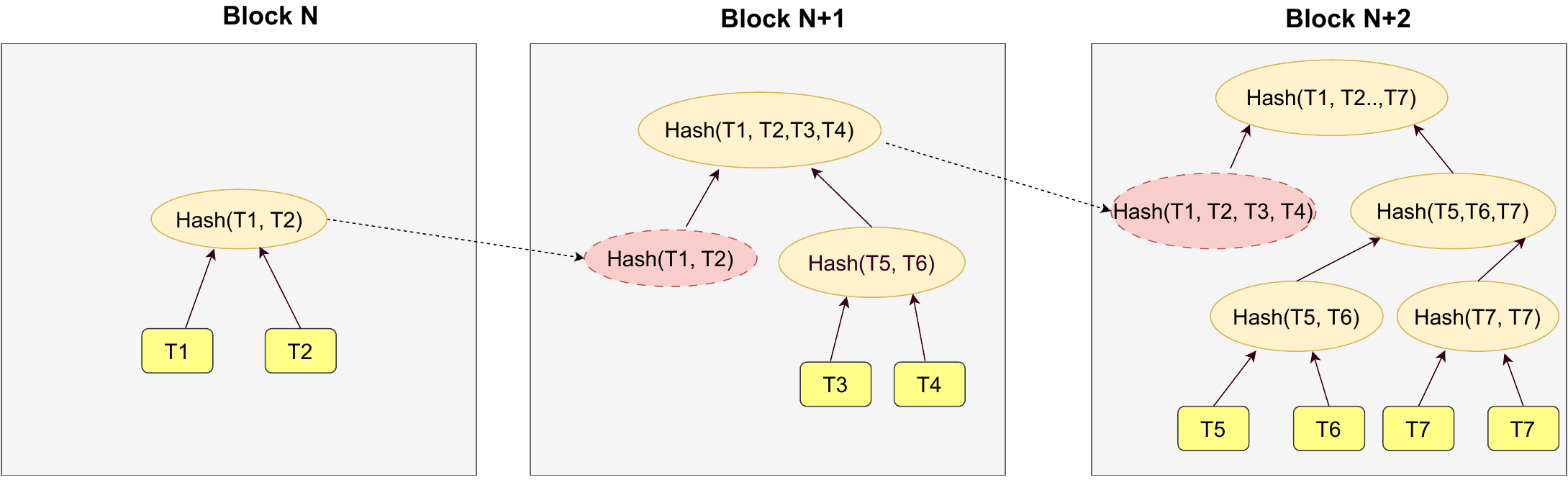}
  \caption{Merkle Root for Distributed Case Tracking}
  \label{fig:case}
\end{figure*}

The second method includes storing the records in an off-chain storage system, allowing faster information retrieval per case. However, when extracting records from the storage, it is essential to verify their trustworthiness. To address this,  we propose that distributed Merkle root for case tracking.
Each case added in the block transactions has its own Merkle root, denoted as $M_{\text{case}}$. As shown in Figure \ref{fig:case}, this Merkle root is calculated as the hash value of the previous Merkle root associated with the case number, hashed with the transactions added in the current block.

Let $M_{\text{prev\_case}}$ denote the previous Merkle root associated with the case number, and let $T_{\text{block}}$ represent the set of transactions added in the current block. The Merkle root for each case, denoted as $M_{\text{case}}$, is calculated as the hash value of the concatenation of $M_{\text{prev\_case}}$ and $T_{\text{block}}$:
\[ M_{\text{case}} = \text{Hash}(M_{\text{prev\_case}} \parallel T_{\text{block}}) \]

This formula ensures that each case added in the block transactions has its unique Merkle root. By hashing the previous Merkle root with the transactions in the current block, we create a tamper-evident structure that links the provenance records of individual cases from one block to the next. Using these two methods, storing provenance records on the blockchain and employing distributed Merkle roots, enhances the system's extraction and verification of provenance records.

%\floatname{algorithm}{Protocol}
\renewcommand{\thealgorithm}{1}
  
    \begin{algorithm}[h]
        \caption{Enhanced Storage Verification}
        \label{algo:storage_verification}
        \begin{algorithmic}
            \Procedure{StorageVerification}{\textit{case\_number}, \textit{sorted\_case\_list}}
                \State $M_{\text{case}} \gets$ null
                
                \For{\textit{block} \textbf{in} \textit{sorted\_case\_record}}
                     \State $M_{\text{block}} \gets$ null
    
                    \State $M_{\text{block}} \gets$ \text{BlockMerkleRoot}(\textit{block\_case\_record})
                    \State $M_{\text{case}} \gets M_{\text{case}} \parallel M_{\text{block}}$
                \EndFor
                
                \State stored\_merkle\_root $\gets$ Retrieve stored Merkle root for the investigation
                
                \If{$M_{\text{case}}$ matches stored\_merkle\_root}
                    \State Data integrity verified
                \Else
                    \State Data integrity compromised
                \EndIf
            \EndProcedure

    \end{algorithmic}
\end{algorithm}

\paragraph{Verification of Provenance records}

\begin{figure}[htbp]
  \centering
  \includegraphics[width=0.5\textwidth]{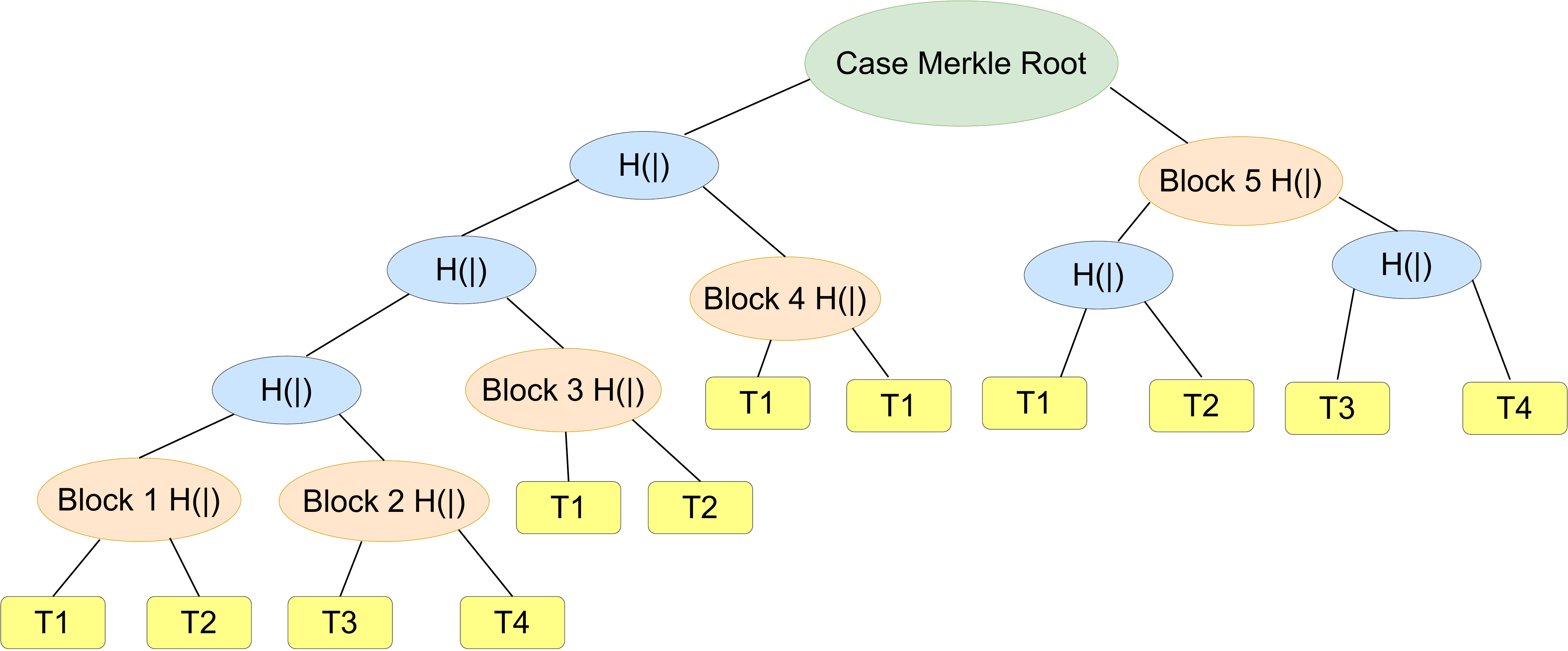}
  \caption{Merkle Tree per each case}
  \label{fig:Merkle-Tree}
\end{figure}

The verification process outlined in Algorithm \ref{algo:storage_verification} is employed to ensure the integrity of storage provenance records per case. This algorithm constructs a Merkle tree, depicted in Figure \ref{fig:Merkle-Tree}, and compares the resulting Merkle root with the stored Merkle root. By performing this comparison, it is determined whether any modifications have occurred in the data.

%% file: Sections/Evaluation.tex
\section{Evaluation}\label{section:Evaluation} 
We developed a prototype of the ForensiBlock system using Python to assess the functionality and performance of the proposed algorithm and protocols \footnote{The GitHub repository link will be shared once permissible upon publication.}. The prototype was tested on a server running the Ubuntu 18:04 TLS operating system, equipped with an Intel(R) Xeon(R) Gold 6140 2.30GHz CPU and 64 GB of RAM. Notably, running the code to replicate the results presented, necessitates only 4GB of memory.
In this section, we conduct an evaluation of the ForensiBlock system, focusing on key aspects such as provenance extraction, smart contract execution, and transaction processing time.

\subsection{Data Generation} 

To evaluate the ForensiBlock system, we conducted experiments that involved manipulating the number of blocks and cases. We generated a comprehensive list of transactions to simulate the activities within the system.
During the transaction generation process, we ensured that each case had an initial transaction known as the \InitalUpload. This transaction represented the start of the case and included a randomly selected stage. In addition to the \InitalUpload, we assigned other transactions to cases by randomly selecting transaction types.
Once the transactions were generated, we sent them to the network for processing. The transactions were then organized and packed into blocks based on the specified number of transactions per block.
By varying the number of blocks and cases in our experiments, we were able to assess the system's performance and behavior under different scenarios.

\subsection{Provenance Extraction}

The objective of this evaluation is to assess the extraction time of provenance records associated with a specific case number within the blockchain. Two different methods are explored for achieving this goal.
The first method involves a brute-force search of the blockchain. It iterates through the blocks until the desired provenance records are found. However, this approach can be computationally intensive and time-consuming.
Alternatively, we can extract the records from the off-chain storage, where the uploaded provenance records are organized based on case numbers. This organized format allows for efficient retrieval of all records associated with a given case number. During the extraction process, the integrity of the retrieved records is validated by verifying the Merkle root of the corresponding case number. This ensures that the provenance information remains intact and untampered.

\begin{figure}[htbp]%\centering
  \centering
  \begin{subfigure}[b]{0.6\textwidth}
    \centering
    \includegraphics[width=0.7\textwidth]{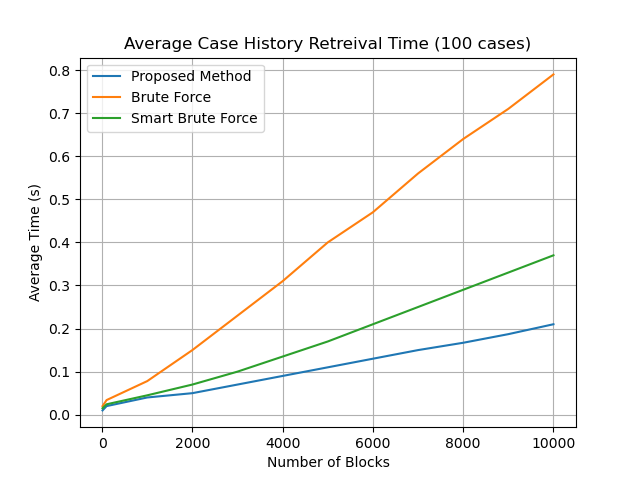}
    \caption{Distribution of 100 cases}
    \label{fig:newplot_100}
  \end{subfigure}
  \hfill
  \begin{subfigure}[b]{0.6\textwidth}
    \centering
    \includegraphics[width=0.7\textwidth]{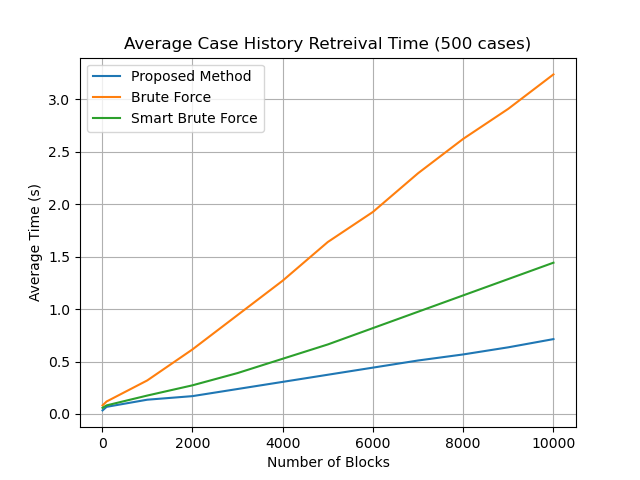}
    \caption{Distribution of 500 cases}
  \label{fig:newplot_500}
  \end{subfigure}
 \hfill
    \begin{subfigure}[b]{0.6\textwidth}
        \centering
        \includegraphics[width=0.7\textwidth]{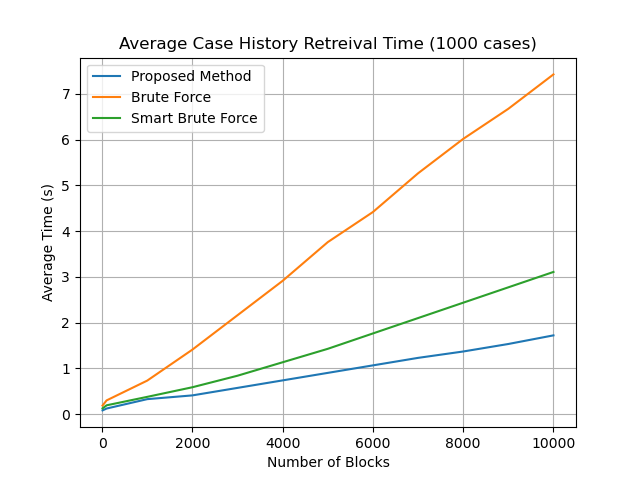}
        \caption{Distribution of 1000 cases}
        \label{fig:newplot_1000}
      \end{subfigure}
      \caption{Average time for retrieving a case provenance records}
      \label{fig:retrive}
\end{figure}

To evaluate the system, we conducted experiments in which we varied the number of blocks from 0 to 10,000, with 100, 500 and 1,000 randomly distributed cases spread across them. The average time to extract case provenance records from the blockchain is presented in Figure \ref{fig:retrive}. The results demonstrate that the proposed method of extracting records from storage and verifying them is unaffected by the size of the blockchain. However, the brute-force method exhibits a significant increase in extraction time as the blockchain grows. Additionally, as the number of cases increases, the retrieval time of records also increases.

%\hoda{what does the following sentence mean? Do you mean The proposed method efficiently verifies the validity of storage records?}
The proposed method efficiently verifies the validity of storage records. However, in cases where the records are invalid, verification with the blockchain itself is necessary. In such scenarios, employing a smart brute-force technique can be beneficial. This technique starts the search from the initial block where a case was created, resulting in faster verification time.
Moreover, the smart brute-force technique exhibits less increase in retrieval time compared to other methods when the number of cases increases.

\subsection{Distributed Merkle Root Creation}
\begin{figure}[htbp]
\centering
\includegraphics[width=0.3\textwidth]{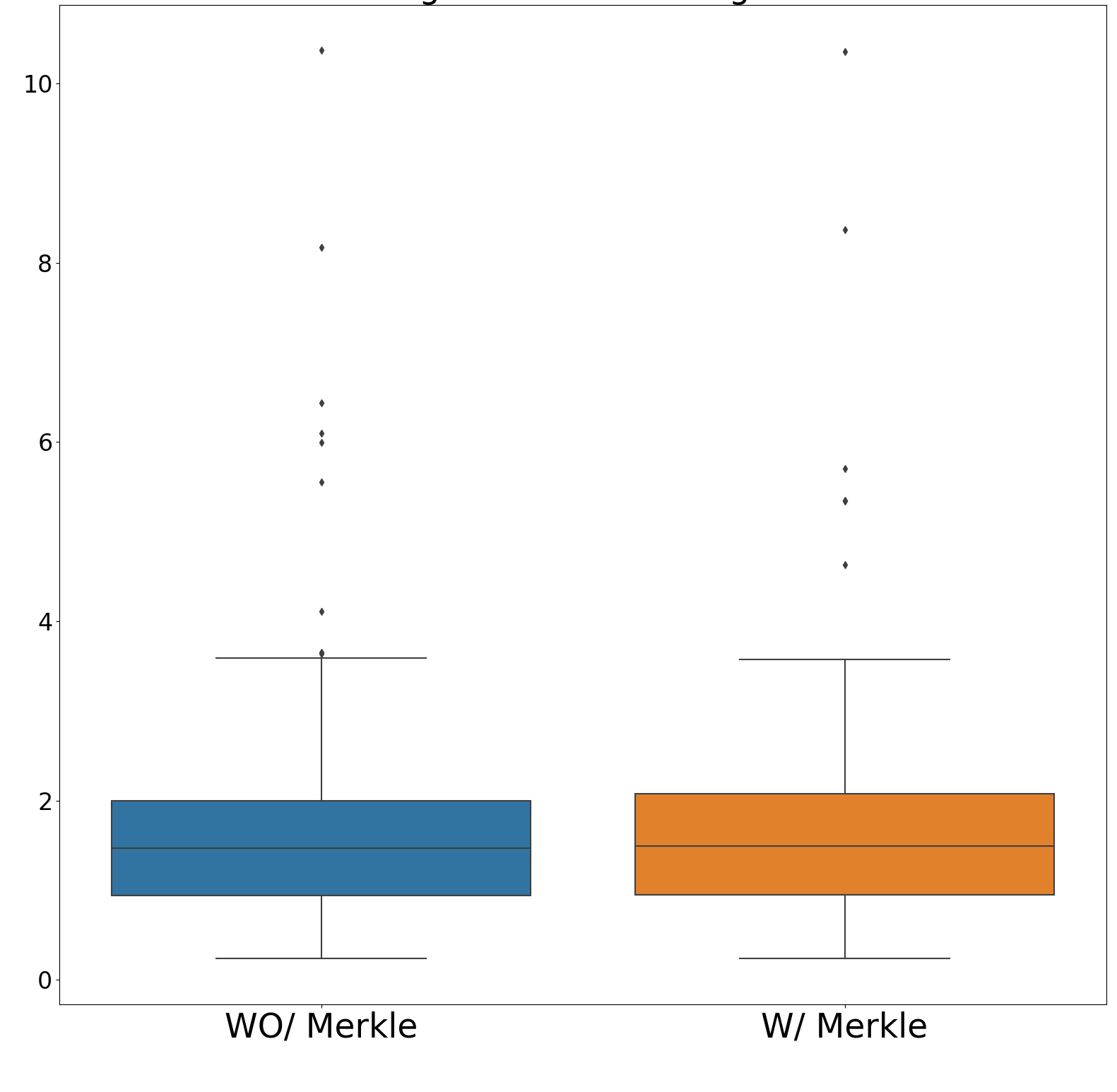}
\caption{Average smart contract processing time}
\label{fig: block processing}
\end{figure}

%In this evaluation, 
We examine the effect of the added features for validating provenance records on smart contract processing and creation time. Figure \ref{fig: block processing} compares the processing time between the provenance system with and without the distributed Merkle root creation.
The box plot illustrates the distribution of data using the lower quartile (Q1), median (m or Q2), upper quartile (Q3), and interquartile range
\begin{equation}
IQR = Q3 - Q1
\end{equation} which represents the central 50\% of the data. The whiskers extend up to 1.5 times the IQR beyond the box. Any data points outside the whiskers are considered outliers and are plotted individually.

Based on the results shown, the range of values for both cases is almost identical. Although our proposed model has a slightly higher average and maximum value (excluding outliers), the difference is negligible, considering the improved history access time.
\subsection{Transaction processing time}
\begin{figure}[htbp]
  \centering
  \begin{subfigure}[b]{0.4\textwidth}
    \centering
    \includegraphics[width=\textwidth]{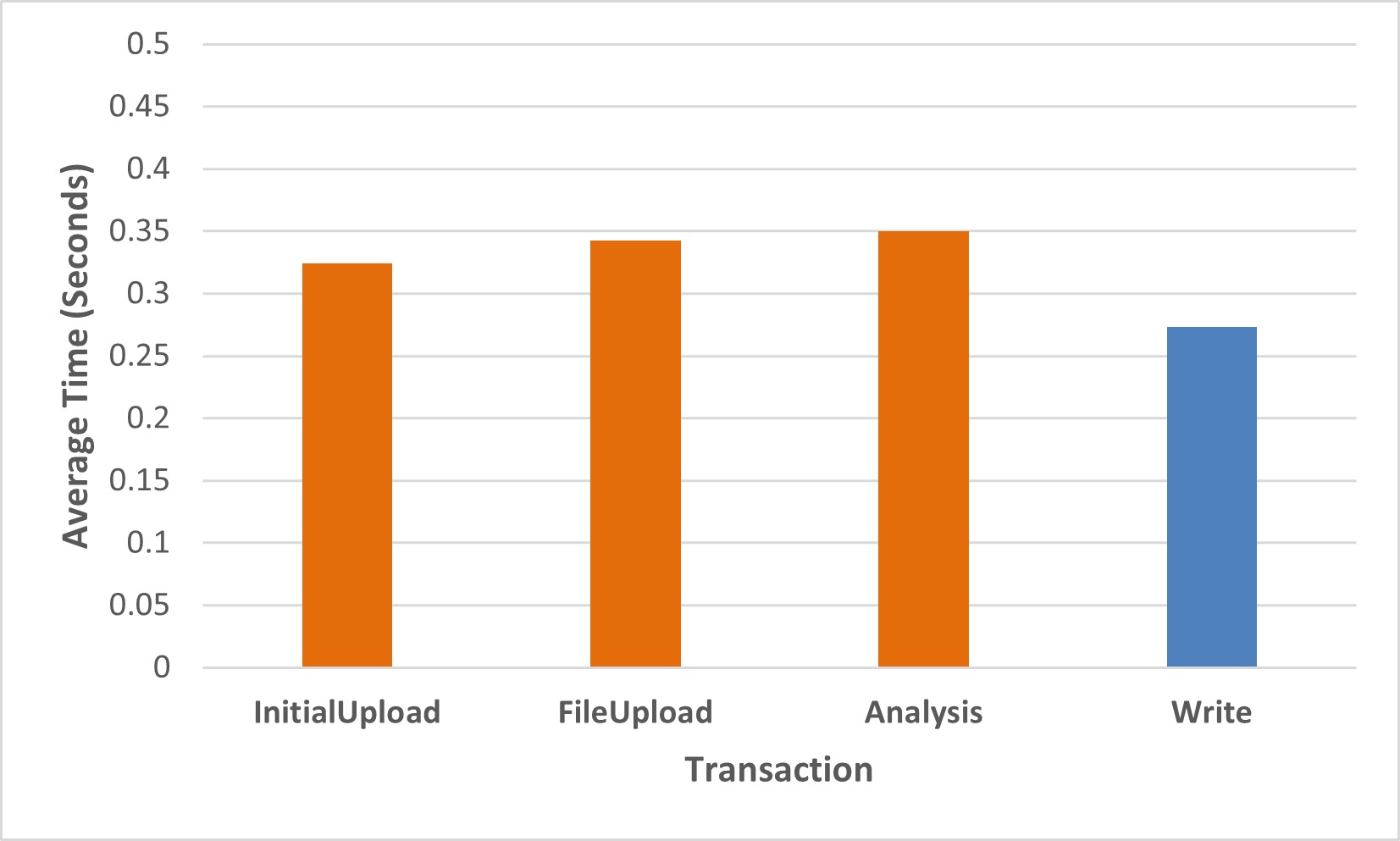}
    \caption{Modifications}
    \label{fig:write2}
  \end{subfigure}
  \hfill
  \begin{subfigure}[b]{0.4\textwidth}
    \centering
    \includegraphics[width=\textwidth]{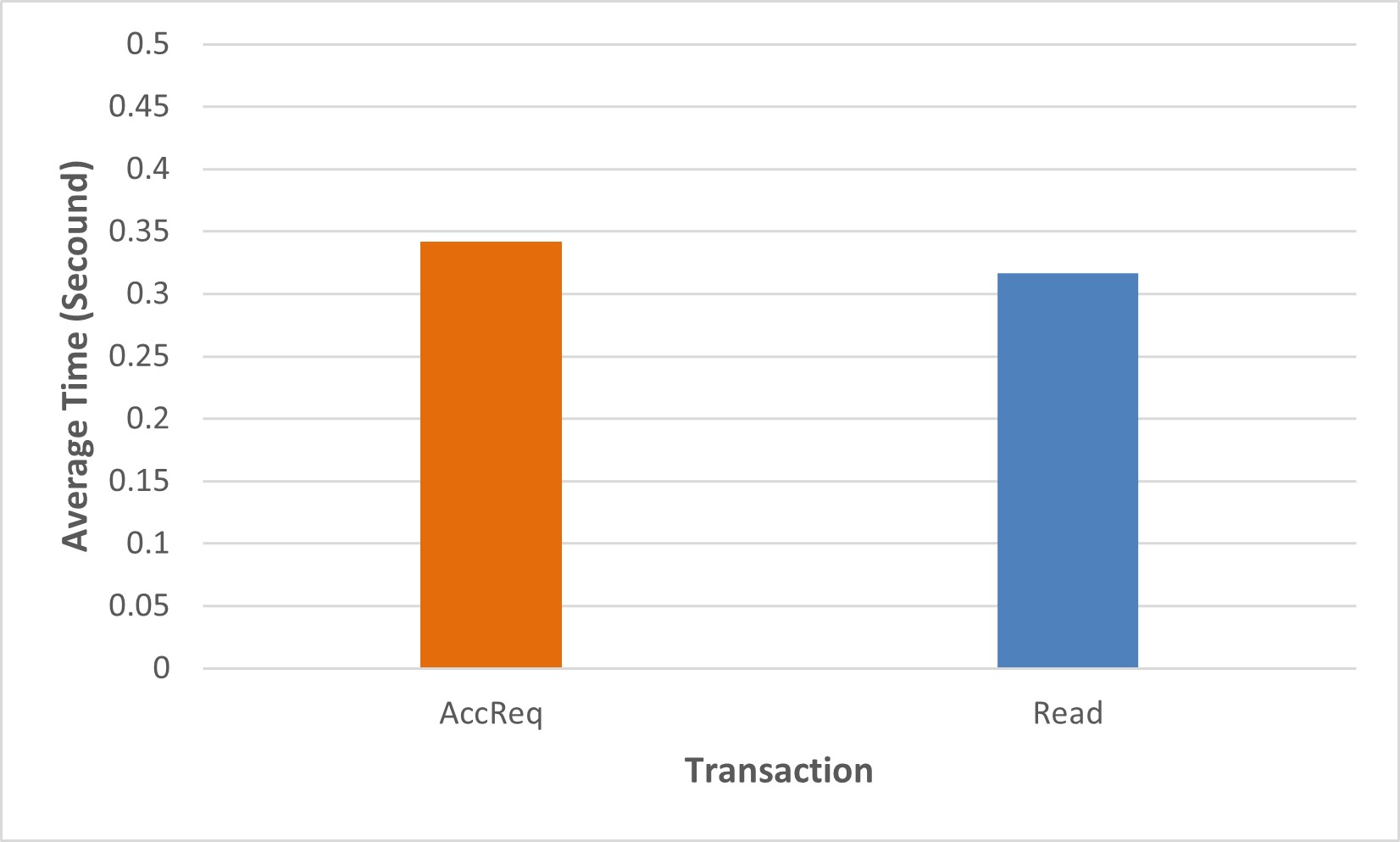}
    \caption{Retrieval}
    \label{fig:read}
  \end{subfigure}
  \caption{Transaction Processing Time}
  \label{fig:transaction_time}
\end{figure}

The transaction processing time in ForensiBlock highlights the distinction between retrieval and modification operations compared to the standard read and write operations used solely for logging purposes. Figure \ref{fig:write2} demonstrates the minor variation between these transaction types, which can be attributed to the specific features they introduce.
For instance, the transactions related to InitialUpload, FileUpload, and Analysis exhibit higher processing times compared to Write transactions due to their involvement in creating new elements or performing more complex operations. On the other hand, the AccReq transaction in \ref{fig:read} surpasses the read operation in terms of processing time due to its role in accessing and adding user rights.

%% file: Sections/Conclusion.tex
\section{Conclusion } \label{section:Con}

 In this paper, we address the importance of provenance in digital forensics and the existing research gap in utilizing blockchain for this purpose. We introduce ForensiBlock, a provenance-driven blockchain solution specifically designed for digital forensics, to overcome the limitations of traditional methods. ForensiBlock ensures comprehensive and transparent record-keeping throughout the investigation process, including steps, extraction, access control, and data version tracking.

The proposed ForensiBlock system incorporates extraction capabilities for timely retrieval of records and a novel access control method to automate digital forensics investigations while ensuring privacy and security. Additionally, we introduce a distributed Merkle tree for verifying the integrity of extracted cases, providing further assurance in the reliability of evidence.

Through the implementation of ForensiBlock and conducting experiments, we demonstrate its capabilities and evaluate its performance. The results highlight the potential of ForensiBlock in enhancing evidence traceability, access control, provenance records, immutability, auditability, fast extraction with verification, and version tracking of data in digital forensics
\section{ACKNOWLEDGMENTS}
This work was funded by NSF grant CCF-2131509

%% file: main.bbl
\begin{thebibliography}{10}
\providecommand{\url}[1]{#1}
\csname url@samestyle\endcsname
\providecommand{\newblock}{\relax}
\providecommand{\bibinfo}[2]{#2}
\providecommand{\BIBentrySTDinterwordspacing}{\spaceskip=0pt\relax}
\providecommand{\BIBentryALTinterwordstretchfactor}{4}
\providecommand{\BIBentryALTinterwordspacing}{\spaceskip=\fontdimen2\font plus
\BIBentryALTinterwordstretchfactor\fontdimen3\font minus
  \fontdimen4\font\relax}
\providecommand{\BIBforeignlanguage}[2]{{%
\expandafter\ifx\csname l@#1\endcsname\relax
\typeout{** WARNING: IEEEtran.bst: No hyphenation pattern has been}%
\typeout{** loaded for the language `#1'. Using the pattern for}%
\typeout{** the default language instead.}%
\else
\language=\csname l@#1\endcsname
\fi
#2}}
\providecommand{\BIBdecl}{\relax}
\BIBdecl

\bibitem{li2019blockchain}
S.~Li, T.~Qin, and G.~Min, ``Blockchain-based digital forensics investigation
  framework in the internet of things and social systems,'' \emph{IEEE
  Transactions on Computational Social Systems}, vol.~6, no.~6, pp. 1433--1441,
  2019.

\bibitem{akhtar2022using}
M.~S. Akhtar and T.~Feng, ``Using blockchain to ensure the integrity of digital
  forensic evidence in an iot environment,'' \emph{EAI Endorsed Transactions on
  Creative Technologies}, vol.~9, no.~31, pp. e2--e2, 2022.

\bibitem{borse2021advantages}
Y.~Borse, D.~Patole, G.~Chawhan, G.~Kukreja, H.~Parekh, and R.~Jain,
  ``Advantages of blockchain in digital forensic evidence management,'' in
  \emph{Proceedings of the 4th International Conference on Advances in Science
  \& Technology (ICAST2021)}, 2021.

\bibitem{chopade2019digital}
M.~Chopade, S.~Khan, U.~Shaikh, and R.~Pawar, ``Digital forensics: Maintaining
  chain of custody using blockchain,'' in \emph{2019 Third International
  conference on I-SMAC (IoT in Social, Mobile, Analytics and
  Cloud)(I-SMAC)}.\hskip 1em plus 0.5em minus 0.4em\relax IEEE, 2019, pp.
  744--747.

\bibitem{pourvahab2019digital}
M.~Pourvahab and G.~Ekbatanifard, ``Digital forensics architecture for evidence
  collection and provenance preservation in iaas cloud environment using sdn
  and blockchain technology,'' \emph{IEEE Access}, vol.~7, pp.
  153\,349--153\,364, 2019.

\bibitem{dasaklis2021sok}
T.~K. Dasaklis, F.~Casino, and C.~Patsakis, ``Sok: Blockchain solutions for
  forensics,'' in \emph{Technology Development for Security
  Practitioners}.\hskip 1em plus 0.5em minus 0.4em\relax Springer, 2021, pp.
  21--40.

\bibitem{liao2021blockchain}
Z.~Liao, X.~Pang, J.~Zhang, B.~Xiong, and J.~Wang, ``Blockchain on security and
  forensics management in edge computing for iot: A comprehensive survey,''
  \emph{IEEE Transactions on Network and Service Management}, vol.~19, no.~2,
  pp. 1159--1175, 2021.

\bibitem{lone2018forensic}
A.~H. Lone and R.~N. Mir, ``Forensic-chain: Ethereum blockchain based digital
  forensics chain of custody,'' \emph{Sci. Pract. Cyber Secur. J}, vol.~1, pp.
  21--27, 2018.

\bibitem{tsai2021application}
F.-C. Tsai, ``The application of blockchain of custody in criminal
  investigation process,'' \emph{Procedia Computer Science}, vol. 192, pp.
  2779--2788, 2021.

\bibitem{ahmed2023using}
M.~Ahmed, A.~R. Pranta, M.~F.~A. Koly, F.~Taher, and M.~A. Khan, ``Using ipfs
  and hyperledger on private blockchain to secure the criminal record system,''
  \emph{European Journal of Information Technologies and Computer Science},
  vol.~3, no.~1, pp. 1--6, 2023.

\bibitem{tasnim2018crab}
M.~A. Tasnim, A.~A. Omar, M.~S. Rahman, and M.~Z.~A. Bhuiyan, ``Crab:
  Blockchain based criminal record management system,'' in \emph{Security,
  Privacy, and Anonymity in Computation, Communication, and Storage: 11th
  International Conference and Satellite Workshops, SpaCCS 2018, Melbourne,
  NSW, Australia, December 11-13, 2018, Proceedings 11}.\hskip 1em plus 0.5em
  minus 0.4em\relax Springer, 2018, pp. 294--303.

\bibitem{bonomi2018b}
S.~Bonomi, M.~Casini, and C.~Ciccotelli, ``B-coc: A blockchain-based chain of
  custody for evidences management in digital forensics,'' \emph{arXiv preprint
  arXiv:1807.10359}, 2018.

\bibitem{zhu2023blockchain}
J.~Zhu, J.~Cao, D.~Saxena, S.~Jiang, and H.~Ferradi, ``Blockchain-empowered
  federated learning: Challenges, solutions, and future directions,'' \emph{ACM
  Computing Surveys}, vol.~55, no.~11, pp. 1--31, 2023.

\bibitem{akbarfam2023dlacb}
A.~J. Akbarfam, S.~Barazandeh, H.~Maleki, and D.~Gupta, ``Dlacb: Deep learning
  based access control using blockchain,'' \emph{arXiv preprint
  arXiv:2303.14758}, 2023.

\bibitem{adhikari2023lockless}
R.~Adhikari and C.~Busch, ``Lockless blockchain sharding with multiversion
  control,'' in \emph{International Colloquium on Structural Information and
  Communication Complexity}.\hskip 1em plus 0.5em minus 0.4em\relax Springer,
  2023, pp. 112--131.

\bibitem{nakamoto2008bitcoin}
S.~Nakamoto, ``Bitcoin: A peer-to-peer electronic cash system,''
  \emph{Decentralized business review}, p. 21260, 2008.

\bibitem{jing2021review}
S.~Jing, X.~Zheng, and Z.~Chen, ``Review and investigation of merkle tree’s
  technical principles and related application fields,'' in \emph{2021
  International Conference on Artificial Intelligence, Big Data and Algorithms
  (CAIBDA)}.\hskip 1em plus 0.5em minus 0.4em\relax IEEE, 2021, pp. 86--90.

\bibitem{bhutta2021survey}
M.~N.~M. Bhutta, A.~A. Khwaja, A.~Nadeem, H.~F. Ahmad, M.~K. Khan, M.~A. Hanif,
  H.~Song, M.~Alshamari, and Y.~Cao, ``A survey on blockchain technology:
  evolution, architecture and security,'' \emph{IEEE Access}, vol.~9, pp.
  61\,048--61\,073, 2021.

\bibitem{amiri2021permissioned}
M.~J. Amiri, D.~Agrawal, and A.~El~Abbadi, ``Permissioned blockchains:
  Properties, techniques and applications,'' in \emph{Proceedings of the 2021
  International Conference on Management of Data}, 2021, pp. 2813--2820.

\bibitem{vu2023blockchain}
N.~Vu, A.~Ghadge, and M.~Bourlakis, ``Blockchain adoption in food supply
  chains: A review and implementation framework,'' \emph{Production Planning \&
  Control}, vol.~34, no.~6, pp. 506--523, 2023.

\bibitem{abiodun2022data}
O.~I. Abiodun, M.~Alawida, A.~E. Omolara, and A.~Alabdulatif, ``Data provenance
  for cloud forensic investigations, security, challenges, solutions and future
  perspectives: A survey,'' \emph{Journal of King Saud University-Computer and
  Information Sciences}, 2022.

\bibitem{pan2023data}
B.~Pan, N.~Stakhanova, and S.~Ray, ``Data provenance in security and privacy,''
  \emph{ACM Computing Surveys}, 2023.

\bibitem{kent2006sp}
K.~Kent, S.~Chevalier, T.~Grance, and H.~Dang, ``Sp 800-86. guide to
  integrating forensic techniques into incident response,'' 2006.

\bibitem{neisse2017blockchain}
R.~Neisse, G.~Steri, and I.~Nai-Fovino, ``A blockchain-based approach for data
  accountability and provenance tracking,'' in \emph{Proceedings of the 12th
  international conference on availability, reliability and security}, 2017,
  pp. 1--10.

\bibitem{cui2019blockchain}
P.~Cui, J.~Dixon, U.~Guin, and D.~Dimase, ``A blockchain-based framework for
  supply chain provenance,'' \emph{IEEE Access}, vol.~7, pp.
  157\,113--157\,125, 2019.

\bibitem{luthi2020distributed}
P.~L{\"u}thi, T.~Gagnaux, and M.~Gygli, ``Distributed ledger for provenance
  tracking of artificial intelligence assets,'' \emph{Privacy and Identity
  Management. Data for Better Living: AI and Privacy: 14th IFIP WG 9.2,
  9.6/11.7, 11.6/SIG 9.2. 2 International Summer School, Windisch, Switzerland,
  August 19--23, 2019, Revised Selected Papers 14}, pp. 411--426, 2020.

\bibitem{zhang2017blockchain}
Y.~Zhang, S.~Wu, B.~Jin, and J.~Du, ``A blockchain-based process provenance for
  cloud forensics,'' in \emph{2017 3rd IEEE international conference on
  computer and communications (ICCC)}.\hskip 1em plus 0.5em minus 0.4em\relax
  IEEE, 2017, pp. 2470--2473.

\bibitem{siddiqui2023activity}
M.~S. Siddiqui, ``Activity provenance for rights and usage control in
  collaborative mr using blockchain,'' in \emph{2023 20th Learning and
  Technology Conference (L\&T)}.\hskip 1em plus 0.5em minus 0.4em\relax IEEE,
  2023, pp. 26--31.

\bibitem{troyer2021privacy}
D.~Troyer, J.~Henry, H.~Maleki, G.~Dorai, B.~Sumner, G.~Agrawal, and J.~Ingram,
  ``Privacy-preserving framework to facilitate shared data access for wearable
  devices,'' in \emph{2021 IEEE International Conference on Big Data (Big
  Data)}.\hskip 1em plus 0.5em minus 0.4em\relax IEEE, 2021, pp. 2583--2592.

\bibitem{hoopes2022sciledger}
R.~Hoopes, H.~Hardy, M.~Long, and G.~G. Dagher, ``Sciledger: A blockchain-based
  scientific workflow provenance and data sharing platform,'' in \emph{2022
  IEEE 8th International Conference on Collaboration and Internet Computing
  (CIC)}.\hskip 1em plus 0.5em minus 0.4em\relax IEEE, 2022, pp. 125--134.

\bibitem{ruan2021lineagechain}
P.~Ruan, T.~T.~A. Dinh, Q.~Lin, M.~Zhang, G.~Chen, and B.~C. Ooi,
  ``Lineagechain: a fine-grained, secure and efficient data provenance system
  for blockchains,'' \emph{The VLDB Journal}, vol.~30, pp. 3--24, 2021.

\bibitem{tosh2019data}
D.~Tosh, S.~Shetty, X.~Liang, C.~Kamhoua, and L.~L. Njilla, ``Data provenance
  in the cloud: A blockchain-based approach,'' \emph{IEEE consumer electronics
  magazine}, vol.~8, no.~4, pp. 38--44, 2019.

\bibitem{demichev2018approach}
A.~Demichev, A.~Kryukov, and N.~Prikhodko, ``The approach to managing
  provenance metadata and data access rights in distributed storage using the
  hyperledger blockchain platform,'' in \emph{2018 Ivannikov Ispras Open
  Conference (ISPRAS)}.\hskip 1em plus 0.5em minus 0.4em\relax IEEE, 2018, pp.
  131--136.

\bibitem{dang2021effective}
T.~K. Dang and T.~A. Duong, ``An effective and elastic blockchain-based
  provenance preserving solution for the open data,'' \emph{International
  Journal of Web Information Systems}, vol.~17, no.~5, pp. 480--515, 2021.

\bibitem{coelho2019blockflow}
R.~Coelho, R.~Braga, J.~M. David, F.~Campos, and V.~Str{\"o}ele, ``Blockflow:
  Trust in scientific provenance data,'' in \emph{Anais do XIII Brazilian
  e-Science Workshop}.\hskip 1em plus 0.5em minus 0.4em\relax SBC, 2019.

\bibitem{ramachandran2018smartprovenance}
A.~Ramachandran and M.~Kantarcioglu, ``Smartprovenance: a distributed,
  blockchain based dataprovenance system,'' in \emph{Proceedings of the Eighth
  ACM Conference on Data and Application Security and Privacy}, 2018, pp.
  35--42.

\bibitem{ramachandran2017using}
A.~Ramachandran and D.~M. Kantarcioglu, ``Using blockchain and smart contracts
  for secure data provenance management,'' \emph{arXiv preprint
  arXiv:1709.10000}, 2017.

\bibitem{nizamuddin2019decentralized}
N.~Nizamuddin, K.~Salah, M.~A. Azad, J.~Arshad, and M.~Rehman, ``Decentralized
  document version control using ethereum blockchain and ipfs,''
  \emph{Computers \& Electrical Engineering}, vol.~76, pp. 183--197, 2019.

\bibitem{fernando2019sciblock}
D.~Fernando, S.~Kulshrestha, J.~D. Herath, N.~Mahadik, Y.~Ma, C.~Bai, P.~Yang,
  G.~Yan, and S.~Lu, ``Sciblock: A blockchain-based tamper-proof non-repudiable
  storage for scientific workflow provenance,'' in \emph{2019 IEEE 5th
  International Conference on Collaboration and Internet Computing
  (CIC)}.\hskip 1em plus 0.5em minus 0.4em\relax IEEE, 2019, pp. 81--90.

\bibitem{wittek2021blockchain}
K.~Wittek, N.~Wittek, J.~Lawton, I.~Dohndorf, A.~Weinert, and A.~Ionita, ``A
  blockchain-based approach to provenance and reproducibility in research
  workflows,'' in \emph{2021 IEEE International Conference on Blockchain and
  Cryptocurrency (ICBC)}.\hskip 1em plus 0.5em minus 0.4em\relax IEEE, 2021,
  pp. 1--6.

\bibitem{al2021scichain}
A.~Al-Mamun, F.~Yan, and D.~Zhao, ``Scichain: Blockchain-enabled lightweight
  and efficient data provenance for reproducible scientific computing,'' in
  \emph{2021 IEEE 37th International Conference on Data Engineering
  (ICDE)}.\hskip 1em plus 0.5em minus 0.4em\relax IEEE, 2021, pp. 1853--1858.

\bibitem{akbarfam2021proposing}
A.~J. Akbarfam and K.~K. Motarjemi, ``Proposing a new protocol for using
  device-to-device communications in narrowband iot-based systems,'' in
  \emph{2021 11th smart grid conference (SGC)}.\hskip 1em plus 0.5em minus
  0.4em\relax IEEE, 2021, pp. 1--5.

\bibitem{hu2020survey}
R.~Hu, Z.~Yan, W.~Ding, and L.~T. Yang, ``A survey on data provenance in iot,''
  \emph{World Wide Web}, vol.~23, pp. 1441--1463, 2020.

\bibitem{caro2018blockchain}
M.~P. Caro, M.~S. Ali, M.~Vecchio, and R.~Giaffreda, ``Blockchain-based
  traceability in agri-food supply chain management: A practical
  implementation,'' in \emph{2018 IoT Vertical and Topical Summit on
  Agriculture-Tuscany (IOT Tuscany)}.\hskip 1em plus 0.5em minus 0.4em\relax
  IEEE, 2018, pp. 1--4.

\bibitem{pahl2018architecture}
C.~Pahl, N.~El~Ioini, S.~Helmer, and B.~Lee, ``An architecture pattern for
  trusted orchestration in iot edge clouds,'' in \emph{2018 third international
  conference on fog and mobile edge computing (FMEC)}.\hskip 1em plus 0.5em
  minus 0.4em\relax IEEE, 2018, pp. 63--70.

\bibitem{javaid2018blockpro}
U.~Javaid, M.~N. Aman, and B.~Sikdar, ``Blockpro: Blockchain based data
  provenance and integrity for secure iot environments,'' in \emph{Proceedings
  of the 1st Workshop on Blockchain-enabled Networked Sensor Systems}, 2018,
  pp. 13--18.

\bibitem{ali2018secure}
S.~Ali, G.~Wang, M.~Z.~A. Bhuiyan, and H.~Jiang, ``Secure data provenance in
  cloud-centric internet of things via blockchain smart contracts,'' in
  \emph{2018 IEEE SmartWorld, Ubiquitous Intelligence \& Computing, Advanced \&
  Trusted Computing, Scalable Computing \& Communications, Cloud \& Big Data
  Computing, Internet of People and Smart City Innovation
  (SmartWorld/SCALCOM/UIC/ATC/CBDCom/IOP/SCI)}.\hskip 1em plus 0.5em minus
  0.4em\relax IEEE, 2018, pp. 991--998.

\bibitem{akbarfamdlacb}
A.~J. Akbarfam, S.~Barazandeh, H.~Maleki, and D.~Gupta, ``Dlacb: Deep learning
  based access control using blockchain.''

\bibitem{sandhu1998role}
R.~S. Sandhu, ``Role-based access control,'' in \emph{Advances in
  computers}.\hskip 1em plus 0.5em minus 0.4em\relax Elsevier, 1998, vol.~46,
  pp. 237--286.

\bibitem{hu2015attribute}
V.~C. Hu, D.~R. Kuhn, D.~F. Ferraiolo, and J.~Voas, ``Attribute-based access
  control,'' \emph{Computer}, vol.~48, no.~2, pp. 85--88, 2015.

\bibitem{afonin2016performance}
S.~A. Afonin, ``Performance evaluation of a rule-based access control
  framework,'' in \emph{2016 39th International Convention on Information and
  Communication Technology, Electronics and Microelectronics (MIPRO)}.\hskip
  1em plus 0.5em minus 0.4em\relax IEEE, 2016, pp. 1414--1418.

\end{thebibliography}
